\input harvmac
\input epsf

%
\let\includefigures=\iftrue
%
%
%
\newfam\black
\input rotate
\input epsf
\noblackbox
%
%
\includefigures
\message{If you do not have epsf.tex (to include figures),}
\message{change the option at the top of the tex file.}
\def\figin{\epsfcheck\figin}\def\figins{\epsfcheck\figins}
\def\epsfcheck{\ifx\epsfbox\UnDeFiNeD
\message{(NO epsf.tex, FIGURES WILL BE IGNORED)}
\gdef\figin##1{\vskip2in}\gdef\figins##1{\hskip.5in}
\else\message{(FIGURES WILL BE INCLUDED)}%
\gdef\figin##1{##1}\gdef\figins##1{##1}\fi}
\def\DefWarn#1{}
\def\N{{\cal N}}
\def\figinsert{\goodbreak\midinsert}
\def\ifig#1#2#3{\DefWarn#1\xdef#1{fig.~\the\figno}
\writedef{#1\leftbracket fig.\noexpand~\the\figno}%
\figinsert\figin{\centerline{#3}}\medskip\centerline{\vbox{\baselineskip12pt
\advance\hsize by -1truein\noindent\footnotefont{\bf
Fig.~\the\figno:} #2}}
\bigskip\endinsert\global\advance\figno by1}
\else
\def\ifig#1#2#3{\xdef#1{fig.~\the\figno}
\writedef{#1\leftbracket fig.\noexpand~\the\figno}%
\global\advance\figno by1} \fi

\def\tilde{\widetilde}

\def\subsubsec#1{\bigskip\noindent{\it #1}}
\def\yboxit#1#2{\vbox{\hrule height #1 \hbox{\vrule width #1
\vbox{#2}\vrule width #1 }\hrule height #1 }}
\def\fillbox#1{\hbox to #1{\vbox to #1{\vfil}\hfil}}
\def\ybox{{\lower 1.3pt \yboxit{0.4pt}{\fillbox{8pt}}\hskip-0.2pt}}

\def\rightarrowbox#1#2{
  \setbox1=\hbox{\kern#1{${ #2}$}\kern#1}
  \,\vbox{\offinterlineskip\hbox to\wd1{\hfil\copy1\hfil}
    \kern 3pt\hbox to\wd1{\rightarrowfill}}}

\def\half{{1\over 2}}
\def\Tr{{{\rm Tr~ }}}

\def\Re{{\rm Re\hskip0.1em}}
\def\Im{{\rm Im\hskip0.1em}}

\def\CM{{\cal M}}
\def\CN{{\cal N}}

\def\tilde{\widetilde}

\def\II{\relax{I\kern-.10em I}}

\def\bar{\overline}

\def\IZ{\relax\ifmmode\mathchoice
{\hbox{\cmss Z\kern-.4em Z}}{\hbox{\cmss Z\kern-.4em Z}}
{\lower.9pt\hbox{\cmsss Z\kern-.4em Z}} {\lower1.2pt\hbox{\cmsss
Z\kern-.4em Z}}\else{\cmss Z\kern-.4em Z}\fi}
\def\IB{\relax{\rm I\kern-.18em B}}
\def\IC{{\relax\hbox{$\inbar\kern-.3em{\rm C}$}}}
\def\ID{\relax{\rm I\kern-.18em D}}
\def\IE{\relax{\rm I\kern-.18em E}}
\def\IF{\relax{\rm I\kern-.18em F}}
\def\IG{\relax\hbox{$\inbar\kern-.3em{\rm G}$}}
\def\IGa{\relax\hbox{${\rm I}\kern-.18em\Gamma$}}
\def\IH{\relax{\rm I\kern-.18em H}}
\def\II{\relax{\rm I\kern-.18em I}}
\def\IK{\relax{\rm I\kern-.18em K}}
\def\IN{\relax{\rm I\kern-.18em N}}
\def\IP{\relax{\rm I\kern-.18em P}}

%
\def\inbar{\,\vrule height1.5ex width.4pt depth0pt}

\font\cmss=cmss10 \font\cmsss=cmss10 at 7pt
\def\IR{\relax{\rm I\kern-.18em R}}

\def\lp10{l_P^{10}}
\def\lp11{l_P^{11}}
\def\R11{R_{11}}


\def\a{\alpha}
\def\b{\beta}

\newbox\tmpbox\setbox\tmpbox\hbox{\abstractfont
}
 \Title{\vbox{\baselineskip12pt\hbox to\wd\tmpbox{\hss
 hep-th/0610080} }}
 {\vbox{\centerline{Strebel Differentials With Integral Lengths}
 \bigskip
 \centerline{And Argyres-Douglas Singularities}
 }}
\smallskip
\centerline{Sujay K. Ashok$^a$, Freddy Cachazo$^a$ and Eleonora Dell'Aquila$^b$}
\smallskip
\bigskip
\centerline{\it $^a$Perimeter Institute for Theoretical Physics}
\centerline{\it Waterloo, Ontario, ON N2L 2Y5, Canada}
\bigskip
\centerline{\it $^b$NHETC, Department of Physics, Rutgers University}
\centerline{\it 136 Frelinghuysen Road, Piscataway NJ 08854, USA}
\bigskip
\vskip 1cm \noindent

\input amssym.tex

Strebel differentials are a special class of quadratic differentials with several applications in string theory. In this note we show that finding Strebel differentials with integral lengths is equivalent to finding generalized Argyres-Douglas singularities in the Coulomb moduli space of a $U(N)$ $\N=2$ gauge theory with massive flavours. Using this relation, we find an efficient technique to solve the problem of factorizing the Seiberg-Witten curve at the Argyres-Douglas singularity. We also comment upon a relation between more general Seiberg-Witten curves and Belyi maps.

\Date{October 2006}
%

\lref\KapustinW{
  A.~Kapustin and E.~Witten,
  ``Electric-magnetic duality and the geometric Langlands program,''
  arXiv:hep-th/0604151.
}

\lref\SeibergWone{
  N.~Seiberg and E.~Witten,
  ``Monopoles, duality and chiral symmetry breaking in N=2 supersymmetric QCD,''
  Nucl.\ Phys.\ B {\bf 431}, 484 (1994)
  [arXiv:hep-th/9408099].
}

\lref\SeibergWtwo{
  N.~Seiberg and E.~Witten,
  ``Electric - magnetic duality, monopole condensation, and confinement in N=2
  supersymmetric Yang-Mills theory,''
  Nucl.\ Phys.\ B {\bf 426}, 19 (1994)
  [Erratum-ibid.\ B {\bf 430}, 485 (1994)]
  [arXiv:hep-th/9407087].
}

\lref\Seiberg{
  N.~Seiberg,
  ``Electric - magnetic duality in supersymmetric nonAbelian gauge theories,''
  Nucl.\ Phys.\ B {\bf 435}, 129 (1995)
  [arXiv:hep-th/9411149].
}

\lref\Maldacena{
  J.~M.~Maldacena,
  ``The large N limit of superconformal field theories and supergravity,''
  Adv.\ Theor.\ Math.\ Phys.\  {\bf 2}, 231 (1998)
  [Int.\ J.\ Theor.\ Phys.\  {\bf 38}, 1113 (1999)]
  [arXiv:hep-th/9711200].
}

\lref\Gopakumarone{
  R.~Gopakumar,
  ``From free fields to AdS,''
  Phys.\ Rev.\ D {\bf 70}, 025009 (2004)
  [arXiv:hep-th/0308184].
}

\lref\Gopakumartwo{
  R.~Gopakumar,
  ``From free fields to AdS. II,''
  Phys.\ Rev.\ D {\bf 70}, 025010 (2004)
  [arXiv:hep-th/0402063].
}

\lref\Gopakumarthree{
  R.~Gopakumar,
  ``From free fields to AdS. III,''
  Phys.\ Rev.\ D {\bf 72}, 066008 (2005)
  [arXiv:hep-th/0504229].
}

\lref\CachazoSWone{
  F.~Cachazo, N.~Seiberg and E.~Witten,
  ``Phases of N = 1 supersymmetric gauge theories and matrices,''
  JHEP {\bf 0302}, 042 (2003)
  [arXiv:hep-th/0301006].
}

\lref\CachazoSWtwo{
  F.~Cachazo, N.~Seiberg and E.~Witten,
  ``Chiral rings and phases of supersymmetric gauge theories,''
  JHEP {\bf 0304}, 018 (2003)
  [arXiv:hep-th/0303207].
}

\lref\kennaway{
  K.~D.~Kennaway and N.~P.~Warner,
  ``Effective superpotentials, geometry and integrable systems,''
  Adv.\ Theor.\ Math.\ Phys.\  {\bf 8}, 141 (2004)
  [arXiv:hep-th/0312077].
}

\lref\janik{
  R.~A.~Janik,
  ``Exact $U(N(c)) \rightarrow U(N(1)) \times U(N(2))$ factorization of Seiberg-Witten curves and N = 1 vacua,''  Phys.\ Rev.\ D {\bf 69}, 085010 (2004)
  [arXiv:hep-th/0311093].
}

\lref\Aharony{
  O.~Aharony, Z.~Komargodski and S.~S.~Razamat,
   ``On the worldsheet theories of strings dual to free large N gauge theories,''
  JHEP {\bf 0605}, 016 (2006)
  [arXiv:hep-th/0602226].
}

\lref\moeller{
 N.~Moeller,
``Closed bosonic string field theory at quartic order,''
JHEP {\bf 0411}, 018 (2004) [arXiv:hep-th/0408067].
}

\lref\mulase{
M.~Mulase and M~. Penkava, "Ribbon Graphs, Quadratic Differentials on Riemann Surfaces, and Algebraic Curves Defined over $\bar Q$,"
[arXiv:math-ph/9811024].
}

\lref\CachazoDSW{
  F.~Cachazo, M.~R.~Douglas, N.~Seiberg and E.~Witten,
  ``Chiral rings and anomalies in supersymmetric gauge theory,''
  JHEP {\bf 0212}, 071 (2002)
  [arXiv:hep-th/0211170].
}

\lref\leila{
Leila Schneps, ``The Grothendieck Theory of Dessins d'Enfants," London Mathematical Society Lecture Note Series, vol 200, 1994.
}

\lref\CachazoIV{
  F.~Cachazo, K.~A.~Intriligator and C.~Vafa,
  ``A large N duality via a geometric transition,''
  Nucl.\ Phys.\ B {\bf 603}, 3 (2001)
  [arXiv:hep-th/0103067].
}

\lref\Strebel{
K. ~Strebel, ``Quadratic Differentials," Springer-Verlag, 1984
}

\lref\Gardiner{ Frederick P.~Gardiner, ``Teichmuller theory and
quadratic differentials," John Wiley and Sons, 1987. }

\lref\SaadiZ{
  M.~Saadi and B.~Zwiebach,
  ``Closed string field theory from polyhedra,''
  Annals Phys.\  {\bf 192}, 213 (1989).
}

\lref\SonodaZ{
  H.~Sonoda and B.~Zwiebach,
   ``Closed string field theory loops with symmetric factorizable quadratic differentials,"
  Nucl.\ Phys.\ B {\bf 331}, 592 (1990).
}

\lref\Zwiebach{
  B.~Zwiebach,
  ``How covariant closed string theory solves a minimal area problem,''
  Commun.\ Math.\ Phys.\  {\bf 136}, 83 (1991).
}

\lref\Belopolsky{
  A.~Belopolsky,
  ``Effective Tachyonic potential in closed string field theory,''
  Nucl.\ Phys.\ B {\bf 448}, 245 (1995)
  [arXiv:hep-th/9412106].
 }

\lref\BelopolskyZ{
  A.~Belopolsky and B.~Zwiebach,
  ``Off-shell closed string amplitudes: Towards a computation of the tachyon potential,''
  Nucl.\ Phys.\ B {\bf 442}, 494 (1995)
  [arXiv:hep-th/9409015].
}

\lref\MoellerW{
  N.~Moeller and P.~West,
  ``Arbitrary four string scattering at high energy and fixed angle,''
  Nucl.\ Phys.\ B {\bf 729}, 1 (2005)
  [arXiv:hep-th/0507152].
}

\lref\Moeller{
  N.~Moeller,
  ``Closed bosonic string field theory at quartic order,''
  JHEP {\bf 0411}, 018 (2004)
  [arXiv:hep-th/0408067].
}

\lref\Belyi{
G~.V~.Belyi, ``On Galois extensions of a maximal cyclotonic field," Math. U.S.S.R. Izvestija {\bf 14} (1980), 247-256 }

\lref\Yaakov{
  I.~Yaakov,
   ``Open and closed string worldsheets from free large N gauge theories with adjoint and fundamental matter,''
  arXiv:hep-th/0607244.
}

\lref\ArgyresPS{
  P.~C.~Argyres, M.~R.~Plesser and A.~D.~Shapere,
  ``The Coulomb phase of N=2 supersymmetric QCD,''
  Phys.\ Rev.\ Lett.\  {\bf 75}, 1699 (1995)
  [arXiv:hep-th/9505100].
}

\lref\ArgyresD{
 P.~C.~Argyres and M.~R.~Douglas,
  ``New phenomena in SU(3) supersymmetric gauge theory,''
  Nucl.\ Phys.\ B {\bf 448}, 93 (1995)
  [arXiv:hep-th/9505062].
}

\lref\CachazoV{
  F.~Cachazo and C.~Vafa,
  ``N = 1 and N = 2 geometry from fluxes,''
  arXiv:hep-th/0206017.
}

\lref\ShapereV{
  A.~D.~Shapere and C.~Vafa,
  ``BPS structure of Argyres-Douglas superconformal theories,''
  arXiv:hep-th/9910182.
}

\lref\G{
A. Grothendieck, ``Esquisse d'un Programme," Preprint 1985.
}

\lref\DavidG{
  J.~R.~David and R.~Gopakumar,
  ``From spacetime to worldsheet: Four point correlators,''
  arXiv:hep-th/0606078.
}

\lref\AshokCD{
S.~K.~Ashok, F.~Cachazo and E.~Dell'Aquila, "Children's drawings from Seiberg-Witten curves," {\it Work in progress.}
}

\lref\GiveonKO{
  A.~Giveon, D.~Kutasov and O.~Pelc,
  ``Holography for non-critical superstrings,''
  JHEP {\bf 9910}, 035 (1999)
  [arXiv:hep-th/9907178].
}

\lref\Pelc{
  O.~Pelc,
  ``Holography, singularities on orbifolds and 4D N = 2 SQCD,''
  JHEP {\bf 0003}, 012 (2000)
  [arXiv:hep-th/0001054].
}


\lref\Moellertwo{
  N.~Moeller,
   ``Closed Bosonic String Field Theory at Quintic Order: Five-Tachyon Contact Term and Dilaton Theorem,''
  arXiv:hep-th/0609209.
}

\lref\Kapustin{
 A.~Kapustin,
   ``The Coulomb branch of N = 1 supersymmetric gauge theory with adjoint  and fundamental matter,''
  Phys.\ Lett.\ B {\bf 398}, 104 (1997)
  [arXiv:hep-th/9611049].
}

\lref\DouglasS{
M.~R.~Douglas and S.~H.~Shenker,
  ``Dynamics of SU(N) supersymmetric gauge theory,''
  Nucl.\ Phys.\ B {\bf 447}, 271 (1995)
  [arXiv:hep-th/9503163].
}

\lref\SeibergF{
 N.~Seiberg,
   ``Adding fundamental matter to 'Chiral rings and anomalies in supersymmetric gauge theory',''
  JHEP {\bf 0301}, 061 (2003)
  [arXiv:hep-th/0212225].
}

\newsec{Introduction}

Supersymmetric gauge theories in four dimensions possess a very rich physical and mathematical structure. A surprising recent example is the connection between twisted $\N=4$ theories and the Langlands program \KapustinW. In the 90's, the Seiberg-Witten solution of $\N=2$ theories and their realization of confinement after breaking to $\N=1$ \refs{\SeibergWone,\SeibergWtwo} was a major breakthrough in showing the control supersymmetry gives over quantum corrections. Seiberg dualities \Seiberg\ in $\N=1$ theories show the power of holomorphy in theories of more phenomenological interest. Also, deep connections, such as the AdS/CFT correspondence \Maldacena, have been found between gauge theories and string theories.

In this paper we explore an unexpected connection between Seiberg-Witten theory
and the theory of quadratic differentials on Riemann surfaces. The quadratic differentials that
will be of interest to us are called Strebel differentials \Strebel.
Given a Riemann surface with $n$ marked points, a Strebel
differential induces on the Riemann surface a metric that makes it
look like $n$ semi-infinite cylinders glued along a graph. This
metric is unique up to a choice of $n$ real numbers that correspond
to the radii of the cylinders.

Strebel differentials naturally establish a bijection between $\CM_{g,n}\times \IR_+^n\,$, the moduli space of Riemann surfaces of genus $g$ with $n$ punctures, with a positive real number associated
to each puncture, and the moduli space of metric ribbon graphs -
ribbon graphs with a length associated to each edge \refs{\Gardiner,\Strebel}. Because of
this property, Strebel differentials have been useful in string
field theory \refs{\SaadiZ, \SonodaZ, \Zwiebach}\ and have played a central role in the calculation
of tachyon amplitudes \refs{\BelopolskyZ, \Belopolsky, \Moeller, \MoellerW, \Moellertwo}. More recently, they have appeared in a proposal by Gopakumar \refs{\Gopakumarone,
\Gopakumartwo, \Gopakumarthree} to relate free field theory ribbon
diagrams to closed string worldsheet correlators. See also \refs{\Aharony, \DavidG, \Yaakov}.

A common roadblock in the different attempts to use Strebel
differentials in physics is the difficulty in constructing such
differentials explicitly. This is because the condition for a
quadratic differential to be Strebel is transcendental and expressed
in terms of elliptic functions. Very few explicit differentials are
known, although some numerical \refs{\Moeller, \Moellertwo}\ and perturbative approaches \DavidG\ have been developed.

In this paper we restrict our study to Strebel differentials with
some integral properties. More explicitly, we impose that the
lengths of the edges of the associated ribbon graphs be integers. This condition
leads to a huge simplification of the problem: it replaces
transcendental equations by polynomial equations. The idea of
restricting to Strebel differentials with this property came from physics; in particular, from supersymmetric gauge theories, where a generating function of chiral operators can naturally be interpreted as an abelian differential on a Riemann surface with integral periods \refs{\CachazoDSW, \CachazoSWone, \CachazoSWtwo}.

At first it might appear that imposing the integrality of the lengths is a very strong constraint and that such differentials are very scarce. This, however, is not the case and in later sections we show that one can approximate any Strebel differential, to any desired degree of accuracy, by one that can be obtained by solving polynomial equations.

Moreover, we will show that constructing Strebel differentials with integer lengths is equivalent to solving for a particular class of Seiberg-Witten curves with Argyres-Douglas singularities \ArgyresD\ in the moduli space of a $U(N)$ $\CN=2$ supersymmetric gauge theory with massive flavours. More explicitly, we replace the transcendental equations by the following factorization problem
\eqn\introw{y_{SW}^2 = P^2(z)+B(z) = Q^3(z)R^2(z)}
where $P(z)$, $B(z)$, $Q(z)$ and $R(z)$ are all polynomials.
Although much simpler than the original transcendental equations, solving \introw\ is still a challenging problem in general. However, it turns out that the relation to Strebel differentials allows us to write a differential equation that reduces the problem, by means of solving linear equations, to only a small set of polynomial equations.

In the case we mostly study, which is the case with four punctures, the number of polynomial equations is always two. These are solved by computing the resultant of the two polynomials. These resultants have interesting factorization properties over $\Bbb{Q}$, connected to the fact that Strebel differentials with integral lengths can also be thought of as the pull back of a meromorphic differential on the sphere by a Belyi map \refs{\Belyi, \mulase}. We will comment only briefly on this point, leaving a detailed discussion to a forthcoming publication \AshokCD.

This paper is organized as follows. In section $2$, we review the definition of Strebel differentials, their essential properties and the equations that need to be solved in order to construct them. As mentioned above, these equations are in general transcendental. We use the relation between Strebel differentials and ribbon graphs to explain how to relate Strebel differentials with integer lengths to those with rational lengths. From this connection we explain how to approximate any Strebel differential with arbitrary real lengths by constructing a related differential with integer lengths. In section $3$, we show how the problem of finding Strebel differentials with integer lengths is purely algebraic and we relate this problem to the factorization of Seiberg-Witten curves with Argyres-Douglas singularities. In section $4$, we show how to use the analytic structure of a Strebel differential to construct a differential equation,  which is the main tool that will allow us to solve the examples treated in Section $5$ and in the appendices.  In section $5$, we consider examples that illustrate several points discussed in the previous sections. The examples considered are the sphere with three and four punctures. In the latter set of examples, after discussing the general problem, we specialize to that of equal residues $(n,n,n,n)$ with $n\in \Bbb{N}$. In section $6$, we discuss how
the relation to Strebel differentials makes its appearance directly in gauge theory and in the string theory realizations of it. In section $7$, we place our analysis in a broader mathematical setting and summarize work in progress that generalizes the correspondence discussed here between Seiberg-Witten curves and critical graphs of Strebel differentials to Grothendieck's theory of dessin d'enfants or ``children's drawings". Some technical details and additional examples are collected in the appendices.

\newsec{Strebel Differentials}

The central figure in this paper is a special class of quadratic
differentials on Riemann surfaces called Strebel differentials. In
order to understand their relation to gauge theories we will exploit
many of their properties, in particular their uniqueness, some
equivalent definitions and their relation to metric ribbon graphs.
Therefore, in this section, we will summarize a few relevant facts in
the theory of Strebel differentials.

\medskip

{\it Definition Of Strebel differential}

\medskip

We are only interested in Strebel differentials defined on a Riemann
sphere $S$ with punctures. A Strebel differential \Strebel\ is a
meromorphic quadratic differential with the following properties:

\medskip

$I.\ \phantom{II}$ It is holomorphic on $S\;\backslash \{p_1,\ldots ,p_n\}\,$,
where the $p_i$'s are the location of the

$\phantom{III.\ }$ punctures.

$II.\ \phantom{I}$ It has only double poles located at $z=p_i$ for $i=1,\ldots, n$.

$III.\ $ The set of open horizontal curves is of measure zero.

\medskip

\noindent
A {\it horizontal curve} is defined as follows. Denote a generic
quadratic differential by $\phi(z)dz^2$ and denote by $\gamma$ a curve on
$S$ parameterized by $t$. Then $\gamma$ is a horizontal curve of
$\phi(z)dz^2$ if $\phi(\gamma(t))(\gamma'(t))^2 \in \Bbb{R}^+$ for
all $t$.

A theorem by Strebel guarantees that a differential satisfying all
three conditions exists. Moreover, if the residues at each pole are
given and are in $\Bbb{R}^+$ then the differential is unique.

The open horizontal trajectories
form the {\it critical graph} of $\phi(z)dz^2\,$: the vertices are the zeroes of
$\phi(z)dz^2$ and the edges are open horizontal curves between two zeroes.
For generic locations of the punctures the differential has simple zeroes and
the vertices of the critical graph are  trivalent. Degenerate
configurations with vertices of higher valency occur at the points in the moduli
space where the differential develops higher order zeroes.

A quadratic differential $\phi(z)\, dz^2$ naturally induces a metric
$|\phi(z)| dzd\bar{z}$ on the Riemann surface. This metric is flat almost everywhere
around each pole of $\phi(z)\, dz^2$, except on the zero measure set of open curves.
The poles are pushed to infinity and the surface looks
like a collection of $n$ semi-infinite cylinders glued along the
critical graph of the Strebel differential, where all the curvature
is concentrated. The residues of the differential determine the
radii of the cylinders.

A Strebel differential can also be completely characterized by
giving all the {\it Strebel lengths}, i.e. the lengths of the edges
of the critical graph measured in the Strebel metric. The following
condition is equivalent to the third condition above and is the one
usually implemented in practice to construct a Strebel differential:

\medskip

IIIb.$\quad$ The lengths
$\int_a^b\sqrt{\phi(z)}dz\,$, with $a$ and $b$ zeroes of the differential, are real.

\medskip

\noindent Note that this condition makes sense, since we have
imposed that the residues of the poles be real as well.

\subsec{Finding Strebel Differentials}

In practice, given the location of the punctures $\{p_1,\ldots
,p_n\}$ and the residues $\{m_1,\ldots ,m_n\}$ one constructs a
differential of the form
\eqn\stepone{\phi(z)dz^2 = {Q(z)\over B(z)^2} dz^2\,,}
where $Q(z)$ and $B(z)$ are polynomials of degree $2n-4$ and $n$ respectively\foot{The degree  of $Q(z)$ coincides with the number of simple zeroes of a generic Strebel differential. In the generic case all vertices are trivalent, so using the appropriate relation between the number of edges and vertices in Euler's formula for a planar graph one finds that, if $n$ is the number of faces, the number of zeroes must be $2n-4\,$.}. Usually, one uses the $SL(2,\IC)$ symmetry to fix three of the poles to be located at $\{0,1,\infty\}$. Taking the remaining poles to be at $p_i$, with $i=2,\ldots, n-2$, and the roots (equivalently, the coefficients) of $Q(z)$ to be unknowns, the Strebel differential takes the form
\eqn\fala{\phi(z)dz^2 = -{m^2_{\infty}\over 4\pi^2}{Q(z)\over
z^2(z-1)^2\, \prod_{j=2}^{n-2}(z-p_j)^2}dz^2 \,.}
Here, we have chosen the overall coefficient so that the residue at
infinity is given by $m_{\infty}$.  By imposing that the
residues, which are calculated as the usual residues of simples
poles of the abelian differential $\sqrt{\phi(z)}dz$, are equal to  $m_i\,$, we  can
fix $n-1$ coefficients of $Q(z)$ by solving linear
equations. Therefore, $n-3$ coefficients are left unknown. Notice that
$y^2=Q(z)$ is a Riemann surface of genus $n-3$ and therefore it has
$2n-6$ independent cycles. This is also the number of independent open trajectories of the differential. Imposing that these $2n-6$
lengths be real completely fixes the $n-3$ complex parameters left
in $Q(z)$ as functions of the locations of the punctures and the
residues. Here is where all the complexity of the problem resides.
In order to impose that the lengths are real one has to find the
roots of $Q(z)$, denoted by $\{z_i\}$ and demand that
\eqn\notri{ {\rm Im}\left( \int_{z_i}^{z_j}\sqrt{\phi(z)}dz \right)
= 0.}
These equations are in general very hard to solve and this is what motivated us to search for simpler equations\foot{A perturbative approach to computing the Strebel differential was initiated in \DavidG.}. Interestingly, we will see that imposing the stronger condition of the lengths being integral simplifies the problem considerably.

\subsec{Example: Sphere With Three Punctures}

As a first example, let us consider the case with three punctures.
This case is simple because, since $n=3$, after fixing the
residues there are no unknown parameters left in $Q(z)$. This is
clear from the fact that the only length to be computed is given by
an integral that can be deformed and evaluated as the sum of the
residues, which are real by definition. Therefore, the lengths are
automatically real. If the residues are chosen to be $\{
m_0,m_1,m_{\infty}\}$, it can be shown that the strebel
differential is completely fixed in terms of the residues to be
\eqn\genu{\phi(z)dz^2 = -{m_{\infty}^2\over 4\pi^2}{\left( z^2 -
\left(1+{m_0^2\over m_{\infty}^2}-{m_1^2\over
m_{\infty}^2}\right)z+{m_0^2\over m_{\infty}^2} \right) \over
z^2(z-1)^2}dz^2.}
If $n>3$ the problem of finding the Strebel differential will also
involve imposing conditions on the lengths \notri\ which is much
harder since it involves solving transcendental equations.

\subsec{Relation To Metric Ribbon Graphs}

It is important to keep in mind the relation
between Strebel differentials and ribbon graphs. This relation will
allows us to learn some nontrivial properties of the Strebel
differentials with integer lengths and the polynomial equations that
give rise to them.

The precise statement, specialized to genus zero, is that Strebel
theory establishes a one-to-one correspondence between
$\CM_{0,n}\times \IR_+^n$ (the {\it decorated moduli space} of a
sphere with $n$ punctures) and the space of metric planar ribbon
graphs, i.e. planar ribbon graphs with a length associated to each
edge.

First note that the decorated moduli space $\CM_{0,n}\times \IR_+^n$
coincides with the moduli space of Strebel differentials on the
sphere with $n$ punctures, since we can interpret the positive real
number associated to each puncture as the residue of the
differential at that point. Strebel's uniqueness theorem guarantees
that, for every choice of residues and location of the punctures,
the differential is unique. Therefore we can rephrase the statement
above as a one-to-one correspondence between the space of Strebel
differentials and the space of metric ribbon graphs.

It is easy to see that the statement is true one way. Given a
Strebel differential we have defined the critical graph as the set
of open horizontal trajectories that connect the zeroes of the
differential, with the zeroes being the vertices of the graph. This
is a genus zero graph that has a positive real number, the distance
between the zeroes measured by the Strebel metric, associated to
each edge. This is the metric ribbon graph.

It is also true that it is possible to associate a unique Strebel
differential to any given planar ribbon graph, but the proof is more
difficult (see \Gardiner).

The critical graph of a Strebel differential contains all
information about the Strebel differential and the associated
Strebel metric. Thinking in terms of the critical graph can be
helpful to intuitively understand the properties of the
differential and we will refer to this later in the discussion. In
figure 1 we show the critical graph of a Strebel differential on a
sphere with four punctures that has the topology of a tetrahedron.
The associated metric ribbon graph can be obtained by thinking about
the (thickened) edges as six ribbons that meet at four trivalent
vertices.\foot{There are more topologies in this case but we
postpone their description to section 6.2.}

\medskip
\centerline{\epsfxsize=0.40\hsize\epsfbox{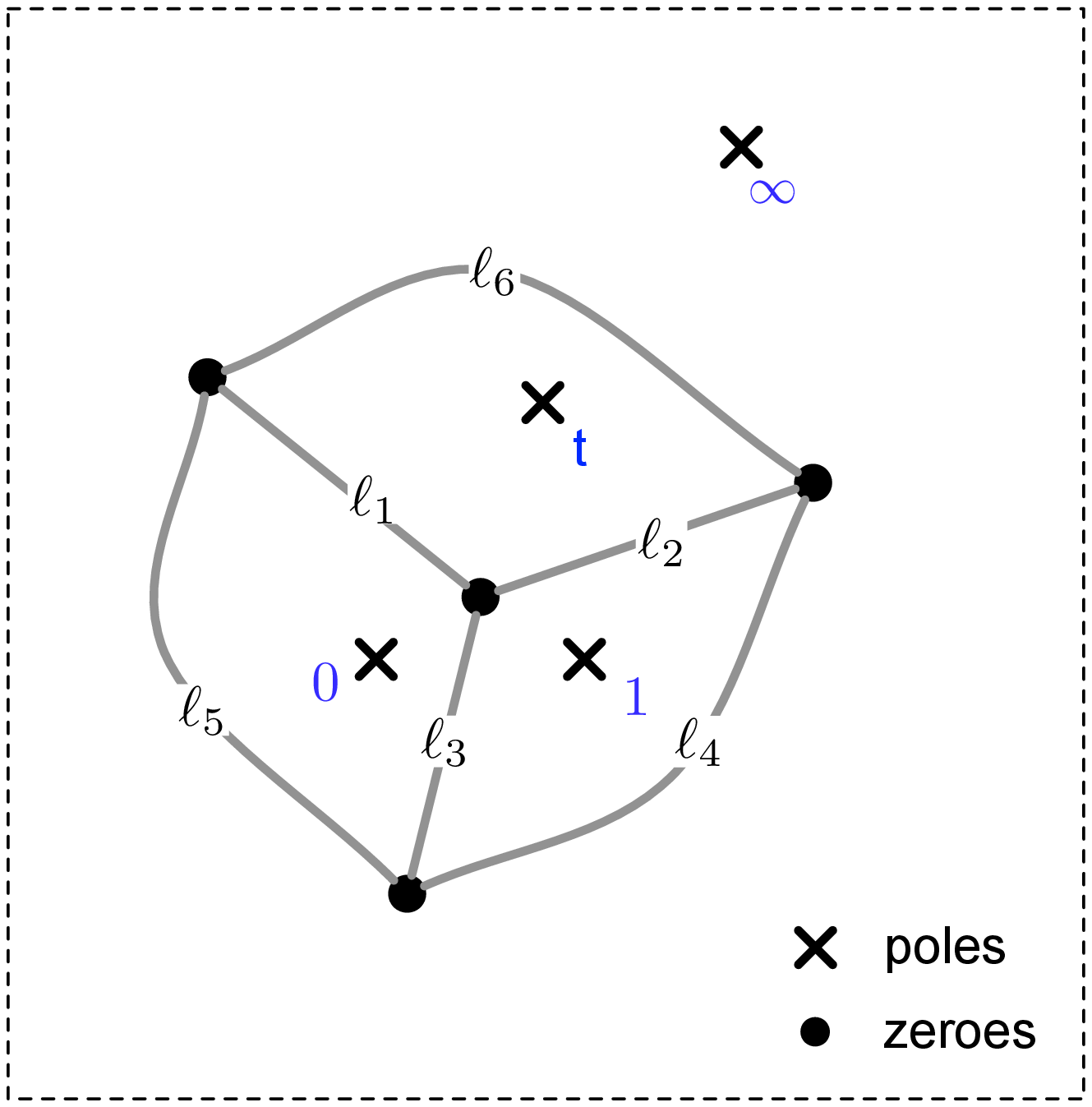}}
\noindent{\ninepoint\sl \baselineskip=3pt {\bf Figure $1$}:{\sl $\;$
Critical graph of a generic Strebel differential on a sphere with
four punctures.}}
\medskip

\subsec{Approximating Any Strebel Differential}

As mentioned in the introduction, in this paper we restrict our
attention to Strebel differentials with integer lengths. Imposing
this condition seems to be a strong constraint; however, from the relation to ribbon
graphs it is clear that this is not so. The condition we are imposing
on the differential is equivalent to asking that the associated
ribbon graph has all integer lengths which can be easily satisfied.

In the rest of this section we give another motivation for
constructing Strebel differentials with integer lengths. We claim
that a generic Strebel differential can be approximated to any
degree of accuracy by a Strebel differential with integer lengths
(divided by some appropriate integer).

The basic idea is the following. Consider a generic Strebel
differential on a sphere with $n$ punctures with an associated
ribbon graph with lengths given by $\{ \ell_1,\ldots
,\ell_{3n-6}\}$.\foot{Recall that the ribbon graph has $n$ faces and
that the perimeter of the face is given by the residue at the
puncture. This is why one usually gives the $n$ residues and $\{
\ell_1, \ldots ,\ell_{2n-6}\}$ independent lengths. However, for this
section it is best to work with all the $3n-6$ lengths and not
mention the residues.} Each $\ell_i\in \Bbb{R}^+$, therefore we can
choose a set of rational numbers that approximate each length to a
prescribed accuracy. In order words, take $\epsilon
> 0 $ and let $|\tilde \ell_i
-\ell_i|<\epsilon\,$, with $\tilde \ell_i\in \Bbb{Q}$. Then let us
consider the approximating Strebel differential $\phi_A(z)dz^2$ with
lengths $\{ \tilde \ell_1,\ldots ,\tilde \ell_{3n-6} \}$. That such a
differential exists is clear from the existence of the corresponding
ribbon graph.

Let $M\in \Bbb{N}$ be the smallest number such that $M\tilde \ell_i \in
\Bbb{N}$ for all $i$. Then $\phi_I(z)dz^2 = M^2\phi_A(z)dz^2$ is a
Strebel differential with integer lengths. Therefore, one can construct
an approximation of the desired differential as
\eqn\guga{\phi_A(z)dz^2 = {1\over M^2}\phi_I(z)dz^2.}

In the next section we will show how to reduce the problem of finding a Strebel differential with integer lengths to a set of algebraic equations and in section $5$ we will study the series of differentials with residues $(n,n,n,n)\,$, with $n\in \Bbb{N}$. From the above discussion, we can think of this series as giving rise to an infinite set of Strebel differentials with residues $(1,1,1,1)$. By explicit computations we will see how the solutions for different values of $n$ correspond to a subset of points in a lattice in ${\cal M}_{0,4}\,$. The lattice gets finer as we increase $n$, so that in the limit $n\rightarrow\infty$, we expect the lattice of points to fill up all of ${\cal M}_{0,4}$.

\newsec{Strebel Differentials With Integral Lengths}

In this section we show that while finding generic Strebel
differentials involves solving transcendental equations, if we
impose that {\it all} the lengths be integers then the problem can
be mapped to a purely algebraic problem. This condition can also be
stated by saying that all independent periods and all residues of
$\sqrt{\phi(z)}dz$, thought of as an abelian differential on
$y^2=Q(z)$, are required to be integers.

The problem of finding an abelian differential on a hyperelliptic
Riemann surface with certain kind of singularities and special
integrality conditions was encountered in
\refs{\CachazoV, \CachazoDSW, \CachazoSWone, \CachazoSWtwo} in the context of $U(N)$ super
Yang-Mills theories. More specifically, in \refs{\CachazoDSW}\ the relevant
abelian differential was the generating function $T(z)dz$ of
$\langle\Tr \Phi^n\rangle$ where $\Phi$ is a chiral supermultiplet
for a $U(N)$ theory with massive flavors in the presence of a tree
level superpotential for $\Phi\,$. We  follow the same steps as in
the construction of \CachazoSWone$\,$, adapted to the particular
singularity structure of our problem. In fact, it will turn out that
the Strebel differential we are after is related very simply to
$T(z)dz$ as
$$2\pi i\, \sqrt{\phi(z)}dz = -2T(z)dz + R(z)dz\,,
$$
for a particular rational function $R(z)\,$.

\subsec{Construction Of The Strebel Differential And Argyres-Douglas Singularities}

We would like an abelian differential on $\Sigma: \; y^2 =Q(z)$ such that all its periods are integers. We can achieve this by taking
\eqn\hulo{\sqrt{\phi(z)}\,dz = {1\over 2\pi i}d \log (f(z))}
for some well-defined meromorphic function $f(z)$ on
$\Sigma$.\foot{Being well-defined on $\Sigma$ means that the only
monodromies allowed around the cycles of $\Sigma$ are $e^{2\pi i
\Bbb{Z}}$.}

We will see below that, given the singularity structure of the
Strebel differential (simple zeroes and double poles), it is more
natural to begin by first constructing a differential of the form
\hulo\ on an ``auxiliary" Riemann surface $\Sigma_0$ defined by the
equation
\eqn\aux{
y_{SW}^2 = P^2(z)+B(z) \,,
}
with $P(z)$ and $B(z)$ polynomials of degrees $N$ and $L$ respectively.

From \hulo\ it is clear that the zeroes or poles of $f(z)$ of degree
$k$ become simple poles of $\sqrt{\phi(z)}\,dz$ with residue $k$ or
$-k$ respectively. Also, since we want $\phi(z)dz^2$ to be well
defined on the sphere, for each pole on the upper sheet there must
be a pole on the lower sheet with the same residue. Since we need
poles on both the lower and the upper sheet of $\Sigma_0\,$, the
function we choose to construct the differential \hulo\ is
\eqn\chofe{ f(z) = {P(z) - y_{SW}\over P(z) + y_{SW}}\,.}
Every zero of $B(z)$ leads to a simple pole in the abelian differential. Likewise, every zero of $y_{SW}$ leads to a zero of the differential. As we will see below, the further requirement that the function $f(z)$ be well-defined on $\Sigma$ leads to constraints that cut down the number of zeroes to exactly the right number for \hulo\ to be the square root of a Strebel differential.

We use an ${\rm SL}(2,\Bbb{C})$ transformation to set three of the $n$ punctures at $\{ 0,1,\infty\}$. There are thus $n-1$ poles at finite points which we identify with zeroes of $B(z)$, each of order $m_i$. More explicitly, if $\{ \infty, 0, 1, p_2,\ldots ,p_{n-2}\}$ are the locations of the poles and $\{ m_\infty, m_0,m_1, m_2,\ldots, m_{n-2}\}$ are the corresponding positive integer residues, then
\eqn\defB{ B(z) = \alpha\,
z^{m_0}(z-1)^{m_1}\prod_{j=2}^{n-2}(z-p_j)^{m_j}\,,}
where at this point $\alpha$ is an arbitrary constant. With this
identification the degree of $B(z)$ is $L=m_0+m_1+m_2+\ldots +
m_{n-2}$. The degree of $P(z)$, denoted by $N$, is related to the
residue at infinity: it is easy to check\foot{If $(L+m_\infty)$ is
not even, we multiply all residues by 2 and divide the differential
by 2.} that $N =(L+m_\infty)/2\,$. For convenience, we choose the
residue at infinity $m_{\infty}$ to be the largest residue.

We would like to identify the curve $\Sigma_0$ with the
Seiberg-Witten curve of an \hbox{${\cal N} =2$} $U(N)$ gauge theory
with $L$ flavors and masses determined by the roots of $B(z)$ (the
locations of the punctures). The curve is not generic: the further
condition that $f(z)$ be well-defined on $\Sigma$ defined by
$y^2=Q(z)$ implies that
\eqn\kino{y_{SW}^2= P^2(z) + B(z) = Q(z)H^2(z)\,,}
where $H(z)$ is another polynomial. Plugging $f(z)$ as given by \chofe\ and \kino\ in the expression \hulo\ for the Strebel differential we get
\eqn\kico{\eqalign{ 2\pi i \,\sqrt{\phi}\, dz &=
{P(z)B'(z)-2B(z)P'(z) \over B(z)\sqrt{P^2(z)+B(z)}}\, dz \cr & =
P(z)Q'(z){H(z)\over \sqrt{Q(z)}B(z)}+
2(-P'(z)H(z)+P(z)H'(z)){\sqrt{Q(z)}\over B(z)}\, dz \,. }}
Comparing with \fala\ we see that this expression has unwanted simple
poles at the zeroes of $Q(z)$, so these
need to cancel for the differential to have the
singularity structure of a Strebel differential. We impose this by requiring that
$Q(z)$ divides $H(z)$, i.e. $H(z)=Q(z)R(z)$ for some polynomial
$R(z)$, and obtain
\eqn\hipo{ 2\pi i \,\sqrt{\phi}\, dz
=(-2P'(z)Q(z)R(z)+3P(z)Q'(z)R(z)+2P(z)Q(z)R'(z))\,{\sqrt{Q(z)}\over
B(z)}\, dz\,.}

Now note that, since \hipo\ was constructed as the logarithmic
derivative of \chofe$\,$, its poles can only be simple
poles at the distinct zeroes of $B(z)$ and a simple pole at infinity
with residue $m_\infty$. Therefore we get the equation
\eqn\diff{ \big[2P'(z)Q(z)R(z)-3PQ'(z)R(z)-2P(z)Q(z)R'(z)\big] =
m_{\infty}\, \prod_{i=0}^{n-2}(z-p_i)^{m_i-1} \,. }
Substituting this into \hipo\ we get the correct expression for the Strebel
differential,
\eqn\jaja{ \sqrt{\phi}\, dz = -{m_{\infty} \over 2\pi i}
{\sqrt{Q(z)} \over \prod_{i=0}^{n-2}(z-p_i)}\, dz \,.}

Note that the Strebel differential obtained this way has integral lengths. Let us see this in more detail. Define $\theta$ to be 
\eqn\deftheta{
\sqrt{\phi(z)}\, dz = {1\over 2\pi i}\, d\, {\rm log}\left({P(z)-y_{SW}\over P(z)+y_{SW}}\right) \equiv {1\over 2\pi}\, d\theta \,.
}
Given $z_a$ and $z_b$ to be any two zeroes of $Q(z)$ (and therefore of $y_{SW}$), we get 
\eqn\strebinteg{
\int_{z_a}^{z_b}\, \sqrt{\phi}\, dz ={1\over 2\pi}( \theta_{b} - \theta_{a}) \in \Bbb{Z} \,,
}
since both $\theta_a$ and $\theta_b$ are equal to $2\pi \Bbb{Z}$. So we have mapped the problem of finding Strebel differentials with integer lengths to that of solving the
algebraic equation
\eqn\qeqi{P^2(z) + B(z) = Q^3(z)R^2(z) \,.}

\subsec{Physical Interpretation}

Let us now comment on the physical interpretation of equation
\qeqi$\,$. It is easiest to start with the curve \kino\
\eqn\alema{y^2 = P^2_N(z) + B_L(z) = Q_{2n-4}(z)H^2_{N-n+2}(z)}
where we have exhibited the degrees of the polynomials explicitly.
This curve arises from a $\N=2$ $U(N)$ gauge theory with $N_f$
massive flavors and tree level superpotential
\eqn\masski{ W_{tree} = \Tr W(\Phi) + \tilde Q_{\tilde f}m_f^{\tilde
f}(\Phi)Q^f}
where $f$ and $\tilde f$ run over the number of flavors $N_f$ and
\eqn\defina{W(z) =\sum_{k=0}^{n-2}{g_k\over k+1}z^{k+1}, \qquad
m_f^{\tilde f}(z) = \sum_{k=1}^{l+1}m_{f, k}^{\tilde f} z^{k-1}.}
The degree of $W(z)$ is $n-1$ and is set by the fact that the
supersymmetric vacua not lifted by \masski\ are those for which
$Q_{2n-4}(z) = W'(z)^2 + f(z)$ where $f(z)$ is a polynomial such
that ${\rm deg}\,(f) ={\rm deg}\,(W'(z))/2 -1$. These turn out to be
$\N=1$ vacua. Finally, $m(z)$ is a matrix of polynomials of size
$N_f\times N_f$. We have denoted the maximum degree of the
polynomials in $m(z)$ by $l$; lower degree ones are obtained by
setting some of the $m_{f,k}^{\tilde f}$'s to zero.

It turns out that the only information about the superpotential
$\tilde Q_{\tilde f}m_f^{\tilde f}(\Phi)Q^f$ which is relevant for
the curve \alema\ is the polynomial \refs{\Kapustin, \SeibergF, \CachazoSWtwo}
\eqn\detti{ B(z)  = {\rm det}\, m(z).}
Clearly, plenty of choices of $m(z)$ can lead to the same $B(z)$.
Recall that we are interested in $B(z)$'s of the form \defB\
\eqn\agapo{ B(z) = \alpha\,
z^{m_0}(z-1)^{m_1}\prod_{j=2}^{n-2}(z-p_j)^{m_j}\,.}
Two natural ways of obtaining such $B(z)$'s are the following:

\item {$\bullet$} $N_f= L={\rm deg} B(z)$, i.e., $N_f = 2N-m_\infty$ and
$m_f^{\tilde f}(z)$ a constant diagonal mass matrix with $m_0$
masses equal to $0$, $m_1$ masses equal to $1$, and $m_j$ masses
equal to $p_j$.

\item {$\bullet$} $N_f= n-1$ and $m_f^{\tilde f}(z)$ a diagonal matrix with
polynomial entries $z^{m_0}$, $(z-1)^{m_1}$, and $(z-p_j)^{m_j}$.

The former leads to a theory with unbroken $\N=2$ supersymmetry if there is
no $W(z)$. Moreover, it has a large flavor symmetry classically. The
latter, on the other hand, has a very small number of flavors and
generically no special flavor symmetry.

Now we are ready to consider \qeqi. What we have done further is to
tune the masses of the flavors (or in the second point of view the
parameters of the polynomials) and the parameters of the
superpotential to arrive at \qeqi$\,$. Near any zero of $Q(z)\,$,
the equation behaves as $y^2=x^3\,$, whose solutions are
Argyres-Douglas points \ArgyresD. These are the points where there are
mutually non-local massless monopoles. Due to the presence of the
superpotential the original monopoles condense and the new
states give rise to $\N=1$ superconformal theories in the IR.

We stress that the factorization problem \qeqi\ is rigid, as it will be easily shown in the
next section. From the point of view of the Strebel differential this means that, for a given set
of integer residues, as we vary the location of the punctures, the
differential has integer lengths only at isolated points in
$\Bbb{C}^{n-3}$. These points are the solutions of \qeqi. In
section $5$ we will study the distribution of these points in the simplest nontrivial case
of $n=4$ and equal residues.

\subsec{Alternative Construction}

Another equivalent way to arrive at \qeqi\ is to start with
\eqn\hulo{\sqrt{\phi(z)}dz = {1\over 2\pi i}d \log \left( {P(z) -
\sqrt{P^2(z) + B(z)}\over P(z) + \sqrt{P^2(z) + B(z)}}\right)}
and realize that near a zero of order $k$ of $P^2(z)+B(z)$ the
abelian differential behaves as $z^{k/2-1}dz$. It then follows that
the corresponding quadratic differential behaves as $z^{k-2}dz^2$.
Therefore, if we only want to have $2n-4$ simple zeroes and nothing
else coming from the $2N$ zeroes of $P^2(z)+B(z)\,$, then $k$ better
be $3$ for $2n-4$ of them and $2$ for the rest. This immediately
leads to \qeqi.

The reason we took the longer route above was to make the relation
to field theory manifest via \kino\ and to lay down the basis for
the construction of a differential equation, which will be the main
tool for finding explicit solutions.

\newsec{Solving Argyres-Douglas Factorizations Using Differential Equations}

For the problem at hand, we have to find some polynomials $P(z)$, $Q(z)$, $B(z)$ and $R(z)$ satisfying the generalized Argyres-Douglas factorization problem \qeqi. We choose $P(z)$, $Q(z)$ and $R(z)$ to be monic. Let us count the number of unknowns and the number of equations. Imposing that $P^2(z)+B(z)$ factors as $Q(z)^3R(z)^2$ is equivalent to requiring $2n-4$ triple roots, which gives $2(2n-4)$ conditions, and $N-3(n-2)$ double roots, which gives $N-3(n-2)$ conditions. The total number of conditions is thus $N+n-2$. Now, the number of unknowns is given by $N$ from $P(z)$, $n-3$ from $p_j$'s in $B(z)\,$ and one more for $\alpha$. This is exactly $N+n-2\,$, justifying our claim that the solutions are isolated points.

Solving such polynomial equations is generically a tedious and  complicated task. In this section we show that, by repeatedly solving linear equations, \qeqi\ can always be reduced to a small number (compared to $2N$) of polynomial equations. We will look in particular at the case with four punctures, for which the number of final equations is two. Since the problem is rigid, i.e. there are only isolated solutions, the two equations are in two variables. A single equation for one of the two variables can then be obtained by computing the resultant of the two polynomials. This final equation factorizes into irreducible polynomials over $\Bbb{Q}\,$, which have an interesting interpretation.

Differentiation tricks are often useful in solving polynomial equations. A simple example is given by the Seiberg-Witten factorization corresponding to a maximally confining point, where $N-1$ mutually local monopoles become massless: $P^2(z)-4=(z^2-4)H^2(z)$. (See appendix A.)

Using the results of the previous section it is possible to find such a differential equation for $P(z)$. Take two expressions for $\phi(z)dz^2$ which are equivalent only when the factorization \qeqi\ holds: for example,  \kico\ and \jaja. By comparing them, we get
\eqn\wewi{ {(P(z)B'(z)-2B(z)P'(z))^2\over B(z)^2(P^2(z)+B(z))} =
{m_\infty^2 Q(z)\over \prod_{i=0}^{n-2}(z-p_i)}\,. }
Substituting in this equation the definition of $B(z)$ from \defB\ we find\foot{This equation can also be derived starting from \diff\ if we use the factorization equation \kino\ and the form of $B(z)$ in \defB.}
\eqn\difftwo{\Big(P(z)\,\sum_{i=0}^{n-2}m_i\prod_{j\ne i}(z-p_j)
-2\,{ dP(z)\over dz}\, \prod_{i=0}^{n-2}(z-p_i)\Big)^2 =
m_{\infty}^2\,Q(z)\Big(P^2(z)+\alpha \prod_{j=0}^{n-2}(z-p_j)^{m_j}
\Big)\,, }
where we have set $p_0=0$ and $p_1=1$ as in previous sections.

Recall from section 2.1 that imposing the residues at the poles
$p_i$ to be equal to $m_i$ fixes $n-1$ out of the $2n-4$ parameters
of $Q(z)$. Moreover, it turns out that the equations to be solved
for the coefficients of $P(z)$ are always linear and can be solved
in terms of the unknown coefficients of $Q(z)$ and $B(z)$. We will
discuss several examples in what follows.

A particularly simple set of residues are those for which
$m_{\infty} > m_0+m_1+\ldots + m_{n-2}\,$. In this case, $P(z)$ can be
found by solving the related differential equation
\eqn\relas{f(z)\,\sum_{i=0}^{n-2}m_i\prod_{j\ne i}(z-p_j) -2\,{
df(z)\over dz}\, \prod_{i=0}^{n-2}(z-p_i) = \sqrt{Q(z)} f(z)}
and taking $P(z) = [f(z)]_+$ (the polynomial part of $f(z)$). The
reason is that when the condition $m_{\infty} > m_0+m_1+\ldots +
m_{n-2}$ is satisfied, the term proportional to $\alpha$ in
\difftwo\ does not affect the highest $N+1$ powers of $z$. Therefore
$P(z)$ can be determined by dropping the $\alpha$ term completely
and taking the square root. This will be used in Appendix $B\,$, where
the series of residues $(1,1,1,3n)$ is solved for any value of $n$
and for a particular location of the pole $p_2$.

\newsec{Examples}

In this section we consider some examples that illustrate how Strebel differentials with integer lengths can be constructed by solving the Argyres-Douglas factorization problem \qeqi. Our main set of examples corresponds to the case of a sphere with four punctures but, as a warm up, we first look at the case with three punctures, which is trivial from the viewpoint of finding explicit Strebel differentials (as shown in section 2.2).

\subsec{Sphere With Three Punctures}

In section 2.2 we gave the explicit form of the Strebel differential
on a sphere with three punctures and residues
$\{m_\infty,m_0,m_1\}$. In this section we want to show what the
corresponding algebraic equation is. The factorization problem is
\eqn\polythree{
P_{N}^2(z) +  B(z) = Q_2^3(z) R_{N-3}^2(z)\,,
}
with
\eqn\juju{ B(z) = \alpha z^{m_0}(z-1)^{m_1}}
and $2N=m_\infty + m_0+m_1\,$ and $L={\rm deg}\, B(z)=m_0+m_1$.

Happily, the solution to the related algebraic equation
\eqn\polyDS{ P_{N}^2(z) +  G_{L}(z) = (1-z^2)\, H_{N-1}^2(z) }
was obtained in \kennaway\ and is given in terms of linear
combinations of Chebyshev polynomials. Here $G_{L}(z)$ is a
polynomial of degree equal to the number of effective flavors in the $U(N)$
theory, as explained in Section $3.2$. In \kennaway\ the masses of all $L$ flavors are arbitrary.

The solution to \polyDS\ is given by
$$\eqalign{
P_N(z) &= \sum_{i=0}^{N_f}\, \nu_i\, \cos((N-i)\theta), \cr
H_{N-1}(z) &=  \sum_{i=0}^{N_f}\,\nu_i {\sin((N-i)\theta)\over \sin
\theta }}$$
where $z= \cos\theta$ and the $\nu_i$'s are constants that will depend on
the masses of the flavors or, equivalently, on the roots of
$G_{L}(z)$.

We still need to impose that
$$
H_{N-1}(z) = Q_2(z) \, R_{N-3}(z)
$$
to get a solution of \polythree. This leads to
two linear equations in the $\nu_i$ variables,
$$
H_{N-1}(-1) = H_{N-1}(1) = 0 \,,
$$
which can be solved easily. The only difficulty arises from requiring $G_{L}(z)$ to  have $m_0$  and $m_1$ coincident roots. The solution obtained in this way can be recast in the original form \polythree\ by an $SL(2,\Bbb{C})$ transformation. This in principle leads to a full solution of the Strebel problem. 

Since the Strebel differential is already known for the case with three punctures and arbitrary residues and the factorization problem is already solved in the gauge theory, we move on to our first nontrivial set of examples.

\subsec{Sphere With Four Punctures}

There are very few explicit examples of non degenerate Strebel differentials
known in the literature. To our knowledge there are only a few
numerical results available for the case with four punctures,
obtained setting all residues equal to one (this case is of interest
for string field theory computations) \moeller\foot{See \Moellertwo\ for very recent numerical results on the five point correlator in string field theory.}. In \moeller\ an
explicit form for the differential is also given, for a very
symmetric point in ${\cal M}_{0,4}$.

Here, after a few comments about the most general case, we will
study in detail the case of equal residues $(n,n,n,n)\,$, with $n\in
\Bbb{N}$.  For a given $n$ we find Strebel differentials with
integer lengths corresponding to all possible partitions of $n$ in
three natural numbers. When we rescale them to reinterpret them as
differentials with residues $(1,1,1,1)$ we find that each solution
(i.e. partition of $n$) corresponds to a different point of ${\cal
M}_{0,4}$. The picture that emerges from the examples that we work
out explicitly is that, if all solutions were considered  for all
$n$, then the set of discrete solutions would fill up ${\cal
M}_{0,4}\,$, as expected from the discussion in section $2\,$.

We will consider more examples in Appendix B.

\subsubsec{General Integral Residues: $\{ m_\infty, m_0, m_1,
m_t\}$}

\bigskip

As explained in Section 3.1, the relevant factorization problem is
\eqn\polyeq{
P_N^2(z) + B(z) = Q_4^3(z)\, R_{N-6}^2(z)\,,
}
with $N={1\over 2}(m_0 + m_1+m_t+m_{\infty})$ and
\eqn\gaga{  B(z) = \alpha z^{m_0}(z-1)^{m_1}(z-t)^{m_t}.}
Using the general form of the Strebel differential in \fala\
and using the fact that the residues at $0,1,t$ are respectively
$m_0,m_1,m_{t}$, we can write $Q_4(z)$ as
\eqn\Qdefn{Q_4(z) = z^4 + q_1 z^3 + q_2 z^2 + q_3 z + q_4\,,}
with $q_2=q_2(q_1,t)$, $q_3=q_3(q_1,t)$ and $q_4=q_4(q_1,t)$ given
by
\eqn\Qdefn{\eqalign{q_2(q_1,t) &= -{m_1^2\over
m_\infty^2}(t-1)+\left({m_0^2\over m_\infty^2}+{m_t^2\over
m_\infty^2}(t-1)\right)t-(1+q_1+t+q_1 t +t^2)\,,\cr 
q_3(q_1,t) & =\left({m_t^2\over m_\infty^2}+(1+q_1)+{m_1^2\over m_\infty^2}
(t-1)+\left(1-{m_t^2\over m_\infty^2}\right)t-{m_0^2\over
m_\infty^2}(1+t)\right)\,t \,,\cr 
q_4(q_1,t) & = {m_0^2\over m_\infty^2} t^2 \,. }}
Therefore, the quadratic differential is specified by the two arbitrary
complex parameters $t$ and $q_1$. Imposing that the differential is
Strebel leads to two (real) independent conditions and thus
determines $q_1$ in terms of $t$. Since we are implicitly imposing
the stronger condition of integer lengths the complex
parameter $t$ is constrained to a discrete set of values, as we will
see in the examples.

Note that even after using the $SL(2,\Bbb{C})$ symmetry to fix the
location of three of the poles of $\phi(z) dz^2$ we still have a
residual symmetry: we can exchange the positions of any two
poles without changing the differential. As a consequence,
$\phi(z)dz^2$ will be symmetric under the group of transformations
generated by $t\to 1-t$ and $t\to -1/t\,$. Moreover, $t\to t^*$ is
also a symmetry of   $\phi\,$. These transformations  have to be
accompanied by corresponding transformations of $q_1$ and $t$ so
that the form \Qdefn\ remains the same. One can check that the
following operations are symmetries:
\eqn\symop{\eqalign{ \{z& \rightarrow 1-z\,, \qquad t \rightarrow 1-t \,,
\qquad q_1 \rightarrow -4-q_1\}\,, \cr
\{ z& \rightarrow {1\over z}\,, \qquad t \rightarrow {1\over t}\,, \qquad q_1 \rightarrow {q_1\over t} \} \,, \cr 
\{ z& \rightarrow z^* \,, \qquad t \rightarrow t^* \,, \qquad q_1 \rightarrow
q_1^* \}\,.\cr }}
For every pair $(t^{(0)},q_1^{(0)})$ one can find five more
solutions to the factorization problem by performing the above operations. To mod out by these symmetries we will restrict $t$ to the fundamental region
bounded by the lines $\Re(t)=\half$, $\Im(t) = 0$ and the curve
$|t|=1$.

\subsubsec{Equal Residues: $(n,n,n,n)$}

\bigskip

We now restrict attention to the case $m_0=m_1= m_\infty = m_t\equiv n\,$. As
mentioned earlier, the simultaneous rescaling of all the residues
does not affect the structure of the differential, therefore we can
use this analysis to get information about the Strebel
differential with residues $(1,1,1,1)\,$ - after re-scaling by $n$ -
for different values of the complex structure parameter $t\,$.
Although the Strebel lengths are integer valued for the $(n,n,n,n)$
case, the rescaled lengths will typically {\it not} be integral.

Let us first discuss some general features of the solutions we will find. Imposing that all the residues are equal leads to a simple form for the Strebel differential:
\eqn\nnnstrebel{ 2\pi\, i\, \sqrt{\phi}\, dz = -n\,{ \sqrt{z^4+q_1
z^3-(q_1(1+t)-2t)z^2+q_1 t z +t^2} \over z(z-1)(z-t)}\, dz \ .}
A Strebel differential on a sphere with four punctures has a
critical graph composed of six edges, with associated Strebel
lengths. Once the four residues are specified, only two of the
lengths are independent. A further simplification occurs when all
residues are equal. The critical graph is a
tetrahedron with all faces of equal perimeter and the lengths of the edges satisfy the relations (with notation that refers
to Figure $1$)
\eqn\indep{ \ell_1 = \ell_4, \qquad \ell_2 = \ell_{5} \qquad
\hbox{and}\qquad \ell_3=\ell_6\,. }
In what follows, we will label the Strebel differentials by three integers corresponding to the lengths $\ell_{1,2,3}$, i.e. the lengths of the edges of a given face of the tetrahedron.

In all the examples we considered we were able to find {\it all} solutions to the factorization problem
\eqn\qoqo{P_N^2(z) + \alpha \left( z(z-1)(z-t)\right)^n =
Q_4^3(z)R^2_{N-6}(z) }
with $N=2n$ and $Q_4(z) =z^4+q_1 z^3-(q_1(1+t)-2t)z^2+q_1 t z +t^2$.
These are given by all possible sets of lengths consistent with
residue $n$. In other words, all possible partitions of $n$ into
{\it exactly} three (strictly positive) integers. The number $p_3(n)$ of such partitions
is given by the generating function
\eqn\case{ {x^3\over (x-1)^3(x+1)(x^2+x+1)} = \sum_{n=0}^{\infty}
p_3(n)x^n\,.}

We obtained these solutions using the differential equation
\difftwo$\,$, which for this case becomes
\eqn\difu{\eqalign{  \Big( n P(z)\,(3z^2-2(1+t)z+t) -2\,{
dP(z)\over dz}&\, z(z-1)(z-t)\Big)^2 =\cr &
n^2\,Q_4(z)\Big(P^2(z)+\alpha \left( z(z-1)(z-t)\right)^n \Big)\,.
}}
We expand both sides of the equation and compare the coefficients of
the various powers of $z\,$. We find that the highest power
$z^{2N+4}$ always gives a trivial condition. Comparing the next $N$
coefficients we find $N$ linear equations for the coefficients of
$P(z)$ (this is always true because of the $P'(z)$ term in the
equation). The next condition, coming from the coefficient of
$z^{N+3}\,$, gives a linear equation for $\alpha$: this is not
obvious, but a careful analysis shows this to be always true. At
this point, all coefficients of $P(z)$ and $\alpha$ are determined
in terms of $q_1$ and $t$.

Taking two more equations, say the coefficients of $z^{N+2}$ and
$z^{N+1}$, one discovers that they are nonlinear polynomial
equations in two variables. The fact that they are nonlinear is
actually good, since we know that we are supposed to find $p_3(n)$
solutions, with $p_3(n)$ given by \case. Taking the resultant of
these two equations we find a single polynomial equation for $t\,$:
let us call it $Res_1(t)$. The equation $Res_1(t)=0$ gives us the
solutions for $t\,$, but it also yields many spurious solutions. The
reason is that there are more coefficients in \difu\ that will
constrain the solution. Taking two more equations, say from the
coefficients of $z^{N}$ and $z^{N-1}$, and computing a second
resultant $Res_2(t)$ one can discard spurious solutions by computing
$$
f(t) = {\rm GCD}\big[Res_1(t), Res_2(t)\big]\,.
$$
The $t$ values that solve the factorization problem are those that
solve  the equation $f(t)=0$. In all our examples this was enough to
discard spurious solutions, but in general one can continue this
process by taking the coefficients of the lower exponents of $z$.

We still need to compute $q_1$. This is done by first computing the resultants $Res_1(q_1)$ and $Res_2(q_1)$ of $f(t)$ with the coefficient of $z^{N+3}$ and $z^{N+2}$ respectively (both of which depend on $q_1$ and $t$). The spurious solutions can be discarded by computing the GCD of these resultants:
$$
g(q_1) = {\rm GCD}\big[Res_1(q_1),Res_2(q_1)\big] \,.
$$
As before, the $q_1$ values that solve the factorization problem are
those that solve the equation $g(q_1)=0$. It turns out that both $f$
and $g$ are of the same degree\foot{This follows from the fact
that each factorization problem has, associated to it, a unique
number field over which the equation factorizes. This statement is
related to the connection with Belyi maps and dessins d'enfants explained in section 7.2.}.

We now turn to study particular cases. The cases $n=1,2$ are not
interesting because from \case\ we see that $p_3(1)=0$ and $p_3(2)=0$.
The way to see this from our factorization problem \qoqo\  is that
the degree of $P_N(z)$ must be at least $N=6\,$. Therefore we need to take $n\ge 3\,$.

\medskip

{$\bullet$ \it $n=3$}

\medskip

For $n=3$ the solution of the factorization problem
$$
P_3^2(z)+\a\, z^3(z-1)^3(z-t)^3 = Q_4^3(z)
$$
is given by
\eqn\soke{t = \half + i{\sqrt{3}\over 2}, \qquad q_1 = -2-i{2\over
\sqrt{3}}\quad \hbox{and}\quad \a = -{64i \over 3\sqrt{3}} }
and the explicit form of the polynomials is
\eqn\solvethree{\eqalign{ P_3(z) &=1+(-1+i\sqrt{3})z+{5\over
2}(1-i\sqrt{3})z^2+{5\over 2}(1+i\sqrt{3})z^4+(-3-i\sqrt{3})z^5+z^6,
\cr Q_4(z)
&=z^4+(-2-{2i\over\sqrt{3}})z^3+(1+i\sqrt{3})z^2-{4i\over\sqrt{3}}z+{1\over
2}(-1+i\sqrt{3})\ .}}
The Strebel differential is
$$
\phi(z)dz^2= -{9\over 4\pi^2 } \,{Q_4(z)\over z^2(z-1)^2(z-{1\over
2 }- i {\sqrt{3}\over 2})^2}\, dz^2 \,.
$$
The zeroes are located at
the vertices of a regular tetrahedron with all the edges of length
one in the Strebel metric. This is the most symmetric point in
${\cal M}_{0,4}\,$, the same found in \moeller$\,$, as mentioned at
the beginning of the section\foot{Our conventions are such that
$-q_1$ maps to the parameter $a$ in \moeller.}.

\medskip
\centerline{\epsfxsize=0.450\hsize\epsfbox{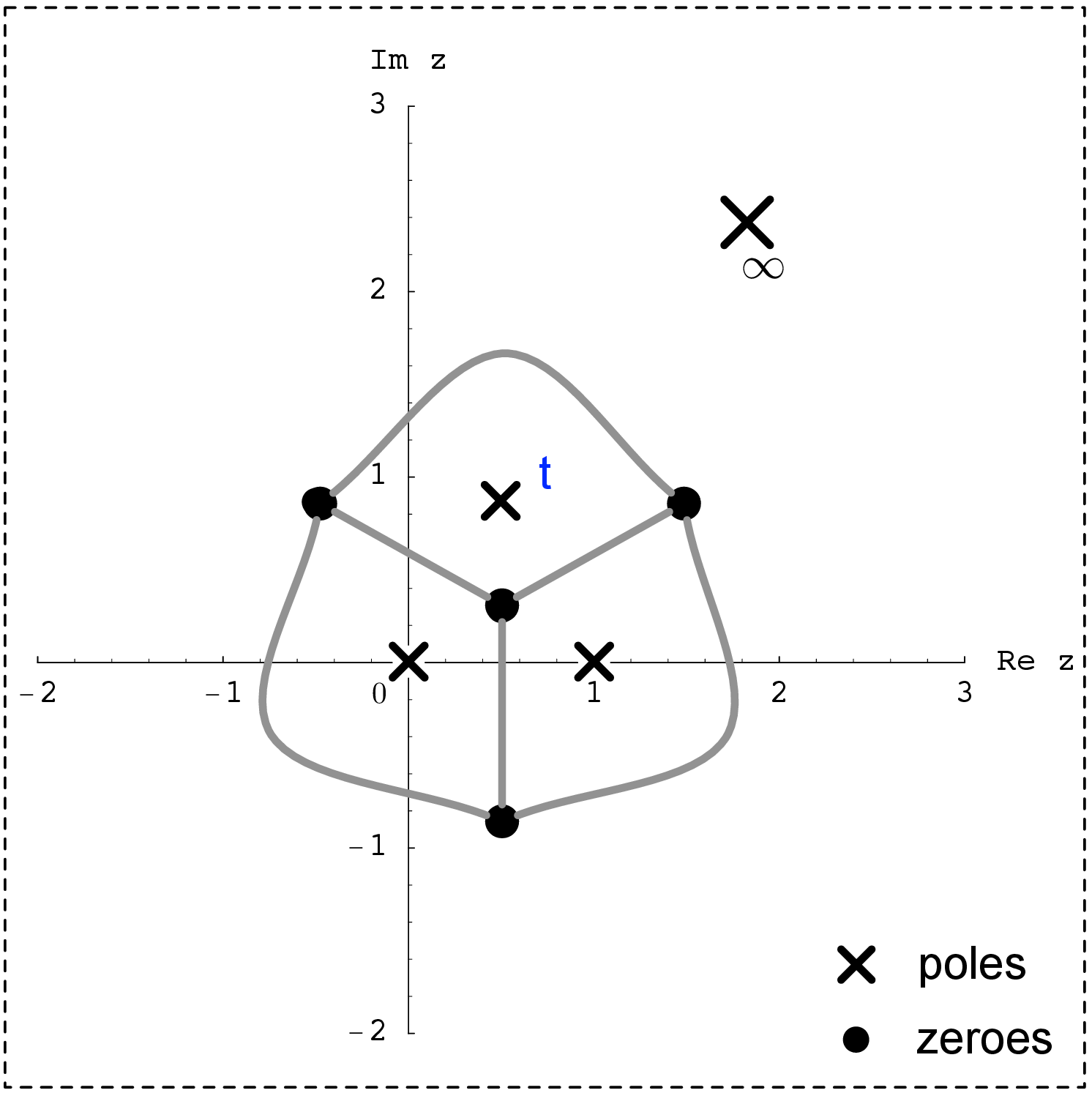}}
\noindent{\ninepoint\sl \baselineskip=3pt {\bf Figure $2$}:{\sl $\;$
Location of poles and zeroes of the Strebel
differential with residues $(3,3,3,3)$.}}
\medskip

{$\bullet$ \it $4\le n\le 9$}

\medskip

For $4\le n\le 9$ the values of $(t,q_1,\alpha)$ and the
corresponding lengths are listed in \hbox{Table $1\,$}. For all the values of
$n$ in the table the form of the polynomials $P(z)$ and $R(z)$ is
easily found, but the expressions get cumbersome. Note that, whenever
two of the three lengths are equal, the location of the third pole
$t$ is fixed to be in the line $\Re(t)=\half$. This automatically
fixes the real part of $q_1$ to be $\Re(q_1)=-2$.

Finally, let us make some remarks about something interesting that
happens when $n$ is not prime. Suppose that $n = rs$ with $r,s>1\,$.
Then from the uniqueness of the Strebel differential it must be that
for  each solution to the problems with $n=r$ or $n=s$ we get a
solution to the problem with $n=rs\,$. This is indeed the case. Take
for example $n=6$ in Table $1\,$: the solution with lengths
$(2,2,2)$ is identical to the one for $n=3$ with lengths $(1,1,1)$.
This observation  seems surprising at first because the
factorization problems for the two cases are quite different.
However,  this is a phenomenon already encountered in the study of
Seiberg-Witten curves and is a consequence of the multiplication map introduced in
\CachazoIV$\,$.

\subsubsec{Relation To Residues $(1,1,1,1)$}

As mentioned earlier, the solutions listed in Table $1$ obtained by choosing integer residues $(n,n,n,n)$ can be recast as solutions to the $(1,1,1,1)$ problem after a suitable re-scaling in each case. In Figure $3$ we have plotted all values of $t$ from Table $1$, interpreting the complex $t$ plane as the moduli space ${\cal M}_{0,4}$ with residues $(1,1,1,1)\,$, a subspace of the full  decorated moduli space ${\cal M}_{0,4}\times \IR^4_+\,$.

In Figure $3\,$, we have used the residual symmetry transformations
\symop\ to bring those points into the fundamental region, whenever
necessary (see section 5.2).  It is clear that one can get an
arbitrarily large number of points in this region by solving
polynomial equations with higher $n$ values. Note that each of
these points is a well defined expansion point, around which a
perturbative analysis similar to that performed in \DavidG\ can be
carried out. It would be interesting to use such an analysis to
find, at least numerically, $q_1(t)$ as a function of the complex
structure parameter $t\,$.

We can see in this example an explicit realization of the general
discussion of Section $2.4\,$, where we showed that it is possible to
approximate any Strebel differential, to arbitrary accuracy, by
solving the problem for integer lengths. Looking at Figure $3$, one
can observe that the points associated to lengths $(1,r,s)$ follow a
regular pattern, forming a grid parametrized by the integers
$(r,s)$. We have tried to make the grid more evident in Figure $4$:
only the points corresponding to the solutions that we have found
explicitly are shown, but one should imagine an infinite grid. The
same observation holds for the points associated to lengths
$(2,r,s)$ and we can presume it to hold more generally for lengths
$(p,r,s)$, for any integer $p\,$.

Note that the lattice $(2,r,s)$ contains the lattice $(1,r,s)$ and
it is finer. This has a simple explanation.  Take the point
corresponding to the lengths $(1,1,2)\,$: this is a node of the
$(1,r,s)$ lattice, but we know that the same value of $t$ is
guaranteed to give a solution for lengths $(2,2,4)$ as well - or any
other multiple. We conclude that the point $(2,2,3)$ of the
$(2,r,s)$ lattice must lie between the origin and the first point of
the $(2,r,s)$ lattice. This simple argument shows how for increasing
$p$ the $(\tilde{p},r,s)$ lattice contains all lattices
corresponding to $p<\tilde{p}$ and it is finer. In the limit $p\to
\infty$ we get a dense set of points covering the fundamental region
of the moduli space.

\centerline{\vbox{\offinterlineskip \hrule \halign{\vrule # &
\strut\ \hfil #\ \hfil & \vrule # & \ \hfil $#$ \hfil \ & \vrule # &
\ \hfil $#$ \hfil \ & \vrule # & \ \hfil $#$ \hfil  \ & \vrule #& \
\hfil $#$ \hfil  \ & \vrule #   \cr height3pt&\omit&&&&&&&&&\cr &n&&
t && q_1 && \a  && \hbox{Lengths}&\cr height3pt&\omit&&&&&&&&&\cr
\noalign{\hrule} height3pt&\omit&&&&&&&&&\cr & 3 && \half+i
{\sqrt{3}\over 2} && -2-i{2\over \sqrt{3}} &&-{64i\over 3\sqrt{3}}&&
(1,1,1) &\cr height3pt&\omit&&&&&&&&&\cr \noalign{\hrule}
\noalign{\hrule} height3pt&\omit&&&&&&&&&\cr & 4 && \half + i
{5\over 2\sqrt{2}} && -2- i\sqrt{2} && -108 && (1,1,2) &\cr
height3pt&\omit&&&&&&&&&\cr \noalign{\hrule} \noalign{\hrule}
height3pt&\omit&&&&&&&&&\cr & 5 &&  \half+i {11\over 6\sqrt{15}} &&
-2-i{2\sqrt{15}\over 9} && {65536 i\over 2025\sqrt{15}}
&&(1,2,2)&\cr height3pt&\omit&&&&&&&&&\cr \noalign{\hrule}
height3pt&\omit&&&&&&&&&\cr &  &&  \half+ i\,2.887948  &&
-2-1.527344 &&1734.654912\, i  &&(1,1,3)&\cr
height3pt&\omit&&&&&&&&&\cr \noalign{\hrule} \noalign{\hrule}
height3pt&\omit&&&&&&&&&\cr &6  &&  \half+i {\sqrt{3}\over 2} &&
-2-i{2\over \sqrt{3}} && {1024\over 27}  &&(2,2,2)&\cr
height3pt&\omit&&&&&&&&&\cr \noalign{\hrule}
height3pt&\omit&&&&&&&&&\cr &  &&  \half+i\, 4.244172
&&-2-i2^{2\over 3} && 43595.583339 &&(1,1,4)&\cr
height3pt&\omit&&&&&&&&&\cr \noalign{\hrule}
height3pt&\omit&&&&&&&&&\cr &  &&-0.127288-i\, 0.622086 &&
-0.625270+i\, 0.793701&& 5.79167+i\, 33.7770&& (1,2,3) &\cr
height3pt&\omit&&&&&&&&&\cr \noalign{\hrule} \noalign{\hrule}
height3pt&\omit&&&&&&&&&\cr &7  &&  \half+i\, 1.292013 && -2-i\,
-1.314895 && -277.399584\, i   &&(2,2,3)&\cr
height3pt&\omit&&&&&&&&&\cr \noalign{\hrule}
height3pt&\omit&&&&&&&&&\cr &  &&0.220923+i\, 0.294611 &&
-1.036986-i\, 0.591846&& 3.207583+i\, 6.143871&& (1,2,4) &\cr
height3pt&\omit&&&&&&&&&\cr \noalign{\hrule}
height3pt&\omit&&&&&&&&&\cr &  &&  \half+i\, 0.341120
&&-2-i\,0.697929 && -7.030850\,i &&(1,3,3)&\cr
height3pt&\omit&&&&&&&&&\cr \noalign{\hrule}
height3pt&\omit&&&&&&&&&\cr &  &&\half+ i\, 5.841390 &&  -2-i\,
1.623197 && -1.569485\times 10^{6} && (1,1,5) &\cr
height3pt&\omit&&&&&&&&&\cr \noalign{\hrule} \noalign{\hrule}
height3pt&\omit&&&&&&&&&\cr &8  &&  \half+i\,{5\over 2\sqrt{2}}  &&
-2-i\,\sqrt{2} && -2916   &&(2,2,4)&\cr height3pt&\omit&&&&&&&&&\cr
\noalign{\hrule} height3pt&\omit&&&&&&&&&\cr &  &&\half+i\, 0.602725
&&  -2-i\, 0.982567&& -23.372595-i\,3.803219&& (2,3,3) &\cr
height3pt&\omit&&&&&&&&&\cr \noalign{\hrule}
height3pt&\omit&&&&&&&&&\cr &  &&\half+ i\, 7.681492&& -2-i\,
1.646274 &&  -7.639365 \times 10^{7}\,i && (1,1,6) &\cr
height3pt&\omit&&&&&&&&&\cr \noalign{\hrule}
height3pt&\omit&&&&&&&&&\cr &  && -0.106193-i\, 0.309236
&&-0.304343+i\, 0.458200 && 6.511956+i\, 18.582505 &&(1,2,5)&\cr
height3pt&\omit&&&&&&&&&\cr \noalign{\hrule}
height3pt&\omit&&&&&&&&&\cr &  &&-0.646518-i\, 1.275342 &&
-0.617371+i\, 0.981889 && -2202.314286-i\, 810.756832 && (1,3,4)
&\cr height3pt&\omit&&&&&&&&&\cr \noalign{\hrule} \noalign{\hrule}
height3pt&\omit&&&&&&&&&\cr &9  &&  \half+i\,{\sqrt{3}\over 2}  &&
-2-i\,{2\over \sqrt{3}} && -{16384\,i \over 811\sqrt{3}}
&&(3,3,3)&\cr height3pt&\omit&&&&&&&&&\cr \noalign{\hrule}
height3pt&\omit&&&&&&&&&\cr &  &&\half+i\, 2.298916 &&  -2-i\,
1.480600&&41259.204415\,i&& (2,2,5) &\cr height3pt&\omit&&&&&&&&&\cr
\noalign{\hrule} height3pt&\omit&&&&&&&&&\cr &  && \half+i\,0.271606
&&-2-i\,0.592769 && 0.000050+i\,6.353074&&(1,4,4)&\cr
height3pt&\omit&&&&&&&&&\cr \noalign{\hrule}
height3pt&\omit&&&&&&&&&\cr &  &&0.945257+0.754441&& -2.965934-i\,
0.988093 &&  -107.573247-i\,19.721375&& (2,3,4) &\cr
height3pt&\omit&&&&&&&&&\cr \noalign{\hrule}
height3pt&\omit&&&&&&&&&\cr &  &&0.123306+\,i0.189777 &&
-0.626671-i\,0.396566 && -4.130947+i\,4.589470 && (1,2,6) &\cr
height3pt&\omit&&&&&&&&&\cr \noalign{\hrule}
height3pt&\omit&&&&&&&&&\cr &  &&\half+i\,9.765341 && -2-i\,1.662029
&&4.824601\times 10^{9}&& (1,1,7) &\cr height3pt&\omit&&&&&&&&&\cr
\noalign{\hrule} height3pt&\omit&&&&&&&&&\cr &
&&1.969844+i\,1.659162 &&-3.384879-i\,1.030218 &&
10737.225688+i\,29575.094989&& (1,3,5) &\cr
height3pt&\omit&&&&&&&&&\cr } \hrule }}
\medskip
\noindent{\ninepoint\sl \baselineskip=3pt {\bf Table $1$}:{\sl $\;$
We list the $t$, $q_1$ and $\a$ values for the $(n,n,n,n)$
problem. The lengths $\ell_1$, $\ell_2$ and $\ell_3$ of Figure $1$ are specified
in the last column:  we get all partitions of $n$ into {\it exactly} three
integers.}}

\medskip
\centerline{\epsfxsize=0.9\hsize\epsfbox{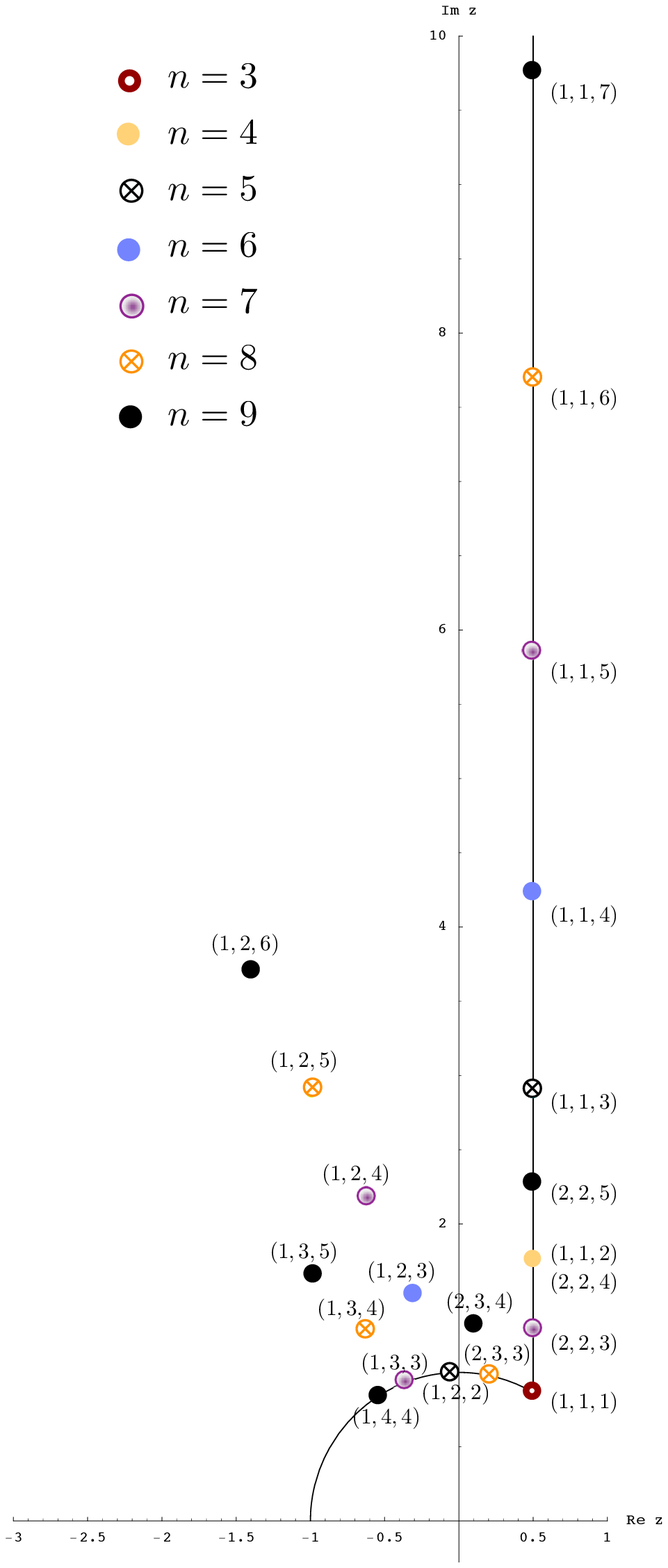}}
\noindent{\ninepoint\sl \baselineskip=3pt {\bf Figure $3$}:{\sl $\;$
Moduli space ${\cal M}_{0,4}$ for residues $(1,1,1,1)$. The points
in the plot are obtained solving the Strebel problem with residues
$(n,n,n,n)\,$. The idea is to rescale by $n$ to obtain a
differential with the desired residues. Next to each point we have
specified the corresponding Strebel lengths $\ell_1,\,\ell_2,\,\ell_3$
(see the discussion around equation \indep).}}
\medskip

\newsec{Relation To Gauge Theory And String Theory}

In this section we expand upon the relation between Strebel differentials and $\CN=2$ gauge theories deformed by a tree level superpotential found in Section $3$. We mention some of the problems on the gauge theory side that can be easily  solved by using the relation to Strebel differentials. We also comment on some intriguing connections to string theory by geometrically engineering the gauge theories. 

\subsec{The Generating Functional Of Chiral Operators $T(z)$}

Strebel differentials with integral lengths, as defined in \hulo\ and \chofe$\,$, are intimately related to the
abelian meromorphic differential $T(z)dz$ defined in \CachazoDSW.
Using
$$
y_{SW}^2 = P^2(z) + B(z)
$$
we get
\eqn\strebelSW{\eqalign{ 2 \pi\, i \sqrt{\phi (z)}\,dz &= d\ln
\left({P(z)-y_{SW}\over P(z)+y_{SW}} \right) \cr &= d\ln B(z)
-2\,d\ln(P(z)+y_{SW}) \cr & = d\ln B(z)-2\,T(z)\,dz \,. \cr }}
Since the first term does not contribute to a period integral
unless it encloses a zero of $B(z)$ (a pole of the Strebel
differential), the factor of $2$ in the relation between $\phi$ and
$T(z)$ ensures that the lengths of the Strebel differential are
identical to the periods of $T(z)\,$.
$$
L_i = {1\over 2\pi i} \oint_{A_i} T(z)\, dz  \equiv N_i \,,
$$
for some choice of $1$-cycles $A_i$ that do not enclose any of the poles. As described in \CachazoDSW, these integers $N_i$ parametrize the classical vacua of a gauge theory in which the gauge group $U(N)$ is broken from $U(N)$ to $\prod_{i=1}^{n} U(N_i)$.

It is important to note that it does not seem possible to interpret the full Strebel differential $\sqrt{\phi(z)}dz$ as the generating functional of chiral operators for a physical theory. On the face of it, it would seem possible that our configuration with poles on the upper and lower sheet can be deformed (by moving the poles through the cuts) into a pseudo confining phase, discussed in \CachazoSWtwo, in which all the poles are on the lower sheet. However, one can check that the starting point itself is unphysical, as some of the period integrals can be shown to be negative. In the gauge theory, these are interpreted as the rank of the 

\medskip
\centerline{\epsfxsize=0.6\hsize\epsfbox{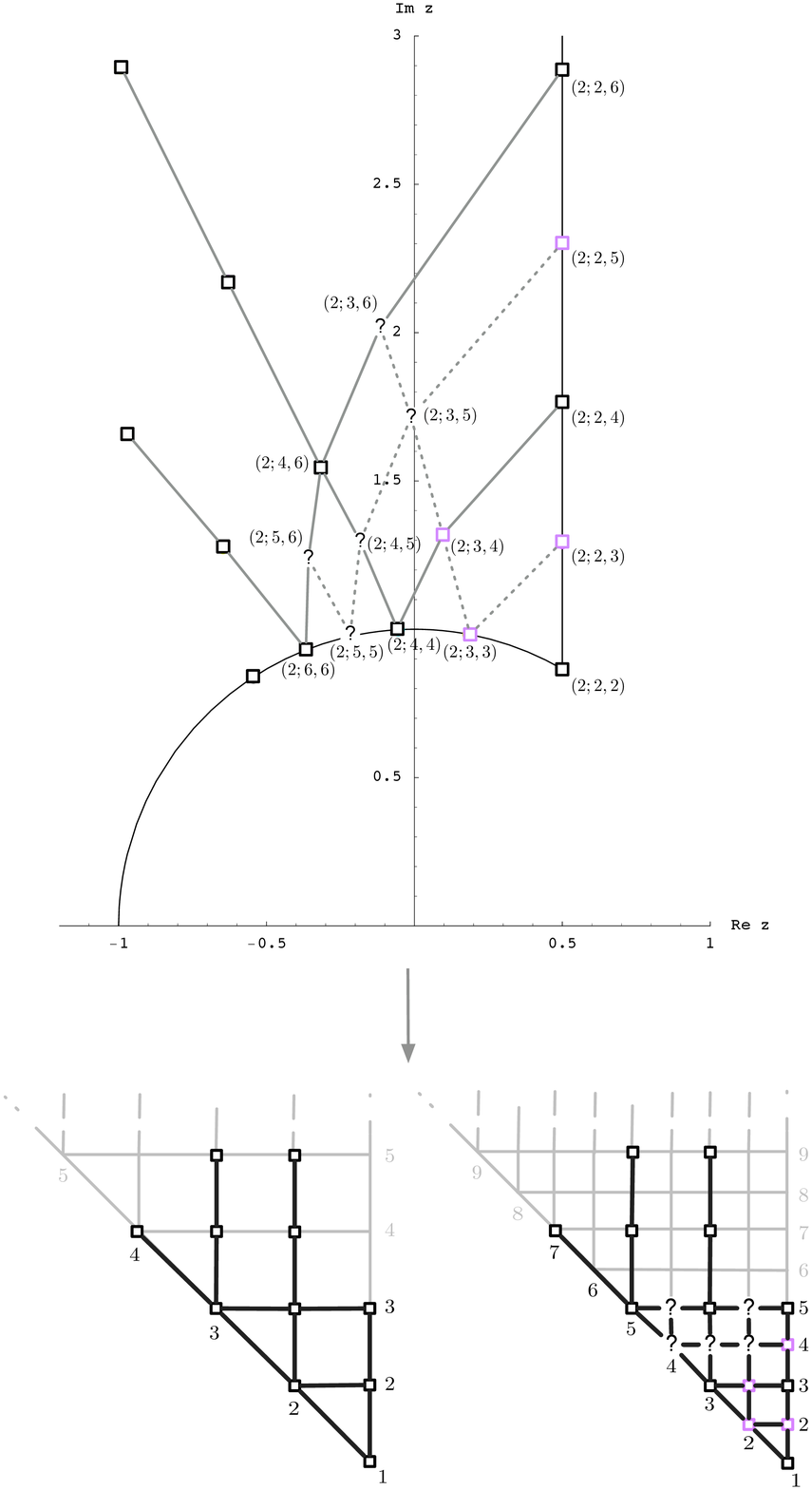}}
\noindent{\ninepoint\sl \baselineskip=3pt {\bf Figure $4$}:{\sl $\;$
On top: Lattice of points with lengths $(2,r,s)$ in ${\cal
M}_{0,4}$. The lattice $(1,r,s)$ is contained in this and it is shown with solid lines. The dashed lines show the refinement of the lattice obtained including all the $(2,r,s)$ points. The points marked by question marks are predictions. Lower left: Embedding in $\Bbb{Z}^2$ of the lattice $(1,r,s)$. Lower right:
Embedding in $\Bbb{Z}^2$ of the lattice $(2,r,s)$. To compare the two lattices we need to shift the coordinates as $(r-p,s-p)$ for the general case of the $(p,r,s)$ lattice. We see clearly from the examples $p=1$ and $p=2$ that increasing $p$ gives a refinement of the lattice.}}
\medskip

\noindent
subgroups $N_i$ into which the original $U(N)$ is broken and so we do not attempt a direct interpretation of this sort.

\subsec{Counting Argyres-Douglas Points}

Consider the gauge theory described in section 3.2. This is a $\N=2$
$U(N)$ gauge theory with matter deformed by a tree-level
superpotential of degree $n-1$ for the adjoint scalar $\Phi\,$ and a
special superpotential $\tilde Q_{\tilde f}m_f^{\tilde f}(\Phi)Q^f$
for the coupling of flavors to $\Phi$.


In section 3.2 we argued that the problem we would like to consider
is the one that leads to a factorization of the form
%
\eqn\jiji{ P^2(z) + B(z) = Q^3(z)H^2(z)}
with $Q(z) = W'(z)^2 + f(z)$ where $f(z)$ is a polynomial of degree
$n-3$ and $W'(z)$ is the derivative of the superpotential function
with degree $n-2$.

At these points in the $\N=2$ moduli space something special
happens: mutually non-local massless particles appear and therefore
the $\N=2$ theories are believed to be superconformal. After adding
the superpotentials, supersymmetry is broken to $\N=1$. The mutually
local monopoles condense and the extra states might lead to $\N=1$
superconformal theories. The physics of this points might be
complicated. However, we would like to concentrate on the problem of
counting such points.


The question is then, how many vacua are there with this property?
%
%

Given just the problem of counting solutions to \jiji, finding the
answer seems very hard. However, from the correspondence to Strebel
differentials and metric ribbon graphs the problem can be easily
converted into a combinatorial problem: to count all possible ribbon
graphs with integer lengths determined by the given residues that
can be drawn on a sphere!

Let us consider some simple examples. Set $n=4$. The factorization
problem is
\eqn\oncemore{P^2_N(z)+\alpha
z^{m_0}(z-1)^{m_1}(z-t)^{m_t}=Q^3(z)R^2(z).}
The problem is to count the number of ribbon graphs with four faces,
only trivalent vertices and such that the perimeter of the faces is
given by $\{ m_\infty, m_0,m_1,m_t \}$ where $m_{\infty} =
2N-m_0-m_1-m_t$. The first step is to determine the possible
topologies of the ribbon graphs. It turns out that in this case
there are only five different topologies. These are shown in Figure $5$. One of them, the
tetrahedron in Figure $5E\,$, was already encountered in section 2.3.

We now choose some particular families of $\{ m_\infty, m_0,m_1,m_t
\}$ as examples.

$\bullet$ $\{ n,n,n,n\}$: This case is by now very familiar. The
only possible topology is a tetrahedron and the number of ribbon
graphs is given by the number of partitions of $n$ into exactly
three integers, i.e.
\eqn\expa{\sum_{n=0}^{\infty} p_3(n)x^n =
x^3+x^4+2x^5+3x^6+4x^7+5x^8+7x^9+\ldots }

$\bullet$ $\{ 1,1,1, 2n+3\}$: There is also only one possible
topology in this case, that of Figure $5A\,$. It consists of three circles, each one connected by a
line to a trivalent vertex. The circles must have circumference one.
The number of possible ribbon graphs is given by the number of
partitions of $n$ in exactly three integers, i.e. $p(n)\,$.

Examples for the other three possible topologies can also be easily
constructed.

\subsec{Relation To String Theory}

The way we have written our Strebel differential is also closely
related to the Seiberg-Witten differential of the corresponding
$\N=2$ theory. For example, looking at equation $(12)$ of
\ArgyresPS\ we see that the SW-differential can be written (after a
suitable shift) as
\eqn\esel{\lambda_{SW}dz = z\ d \ln \left( {P(z)-y_{SW}(z)\over
P(z)+y_{SW}(z)} \right) \,,}
where $y_{SW}^2(z) = P^2(z) + B(z)\,$.

From the work of \CachazoV\ we know that, geometrically engineering this theory in a type IIB superstring theory on a non-compact Calabi-Yau three-fold that undergoes a geometric transition, one finds that the SW-differential is given by
\eqn\juli{ \lambda_{SW}dz = z\ h(z)dz \quad {\rm with} \quad h(z)dz =
\int_{S^2}H_3 \,,}
where $H_3$ is the type IIB three-form field strength $H_{NS}+\tau H_{RR}$ and $\tau$ is  the complexified string coupling. This means that for the geometries realizing the Argyres-Douglas singularities considered in Section $3$, $h(z)dz$ becomes a Strebel differential!

\medskip
\centerline{\epsfxsize=0.8\hsize\epsfbox{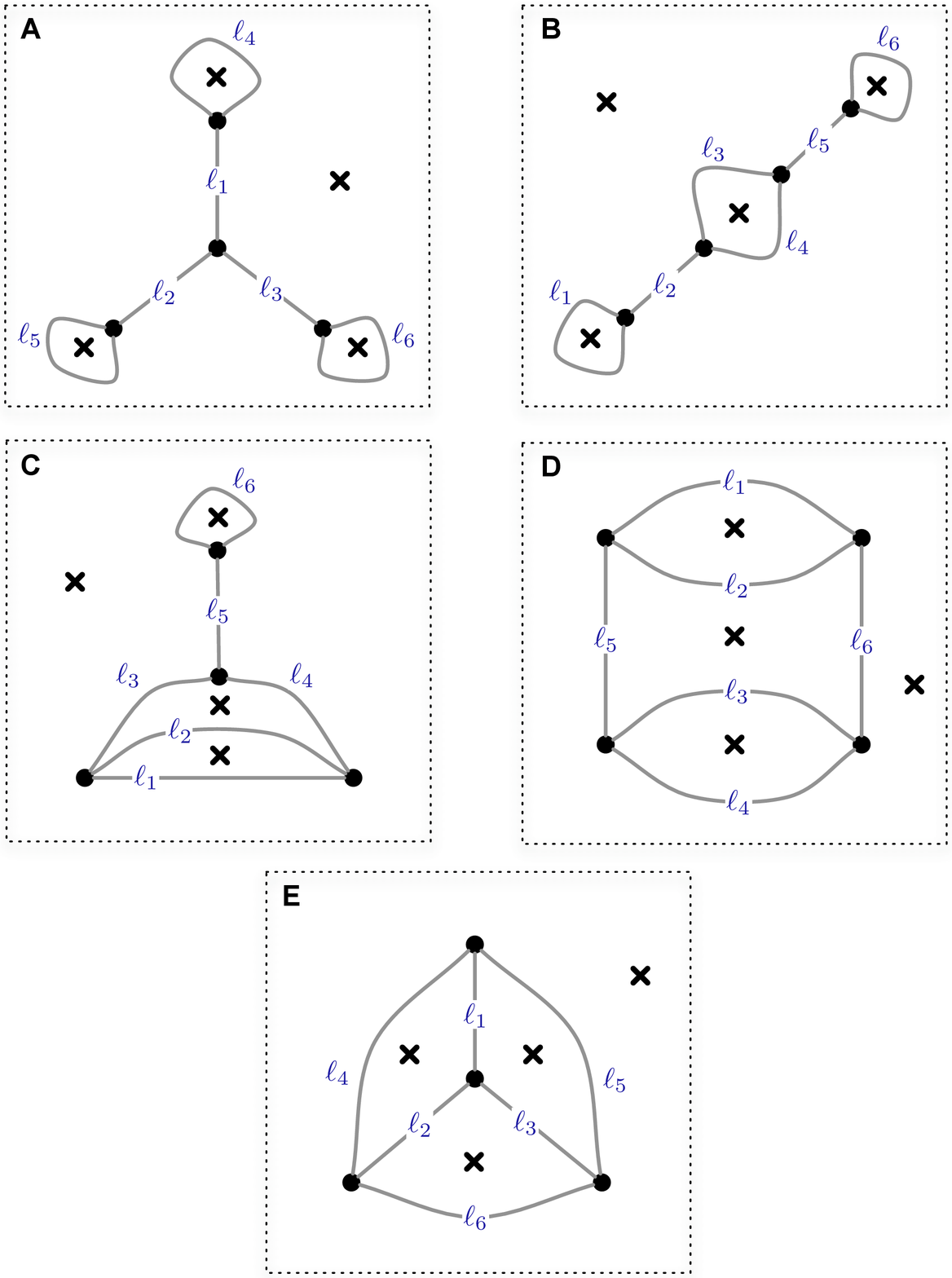}}
\noindent{\ninepoint\sl \baselineskip=3pt {\bf Figure $5$}:{\sl $\;$
The five topologies of the ribbon graphs with four faces.}}
\medskip

This fact might have some connection with the counting of BPS states performed in \ShapereV. In that case finding BPS dyons in Argyres-Douglas superconformal theories was
equivalent to finding special lagrangian cycles, whose definition
contains the condition ${\rm Im}(e^{i\alpha}\Omega) = 0\,$, for some
fixed real $\alpha$ and with $\Omega$ being the holomorphic three
form of the Calabi-Yau three-fold. This is similar to the condition
for a horizontal trajectory discussed in Section $2\,$. Thus the
number of BPS states is roughly speaking the number of horizontal
trajectories. It is interesting to note that the condition for being
Strebel could be interpreted as the condition to have a finite
number of open horizontal trajectories and hence a finite number of
BPS states. An important point to note is that the analysis of
Shapere and Vafa is carried out by first zooming into the
Argyres-Douglas point, which in our case would correspond to the
$SU(3)$ AD curve $y^2 = x^3$, and then counting BPS states after
deforming away from the AD point, by turning on relevant
deformations so that the Seiberg-Witten curve becomes
$y^2=x^3+ux+v\,$. By contrast, our differential is Strebel only at
the AD-point and we retain information about the bigger $U(N_c)$
gauge theory in which the AD point is embedded. Therefore it is not
straightforward to identify objects between the two set-ups. It
would be interesting to clarify the precise relationship between the
two discussions.

\newsec{Strebel Differentials From Belyi Maps}

A surprising property of Strebel differentials was discussed in
\mulase$\,$: it was shown that a Strebel differential with integral
lengths can always be constructed as the pullback by a rational
function $\b(z)$ of a meromorphic quadratic differential on a sphere
with only three punctures: \eqn\pullback{ \phi(z)(dz)^2  =
\b^*\left({1\over 4\pi^2}{d\zeta^2 \over \zeta(1-\zeta) }\right)\ .
} Here the map
$$\eqalign{
\b:\ \Sigma &\rightarrow \IP^1\cr z\,&\mapsto \zeta }$$ must be a Belyi map \Belyi, which means that it must satisfy the property of having {\it exactly} three critical values, at $\{0,1,\infty\}$. Let us see how this comes about in our construction.  

Consider the change of variables
\eqn\zetatheta{
\zeta = \half(1+\cos\theta)\,.
}
After this change of variables the differential on the r.h.s. of \pullback\ takes the form
\eqn\phitheta{
\phi(z)dz^2 ={1\over 4\pi^2} d\theta^2 (z)\, .
}
If we equate this to our expression for the Strebel differential in \hulo\ we get
\eqn\thetaz{
e^{i\theta(z)}  = {P(z) - \sqrt{P^2(z) + B(z)} \over P(z)+ \sqrt{P^2(z) + B(z)}}\,.
}
Note that this is exactly the change of variables performed in \deftheta. Substituting this in the expression for \zetatheta, we get an explicit expression for our candidate Belyi map in terms of the
polynomials that solve the factorization problem \polyeq :
\eqn\belyimap{
\b(z) = {Q^3(z)\, R^2(z) \over \, B(z)} = 1 + {P^2(z)\over B(z)}\,.
}

Let us now show that the map \belyimap\ satisfies the right properties:

$\bullet$ From the two equalities, it is clear that the critical points of $\b$ are the zeroes of $Q(z)$, $R(z)$ and $P(z)$.

$\bullet$ The critical value at the zero of $Q(z)$ or $R(z)$ is zero, while the critical value at a zero of $P(z)$ is equal to $1$.

$\bullet$ $\infty$ is a ramification point on $\Bbb{P}^1$ and the pre-image of $\infty$ is given by zeroes of $B(z)$.
The map \belyimap\ is ramified on $\Bbb{P}^1$ at exactly the points $\{0,1,\infty\}$ and is thus a Belyi map.

\subsec{Example : $(2,2,2)$}

As an example, let us apply this formula to the simple case of the sphere with three punctures, choosing residues $(2,2,2)$. This case is discussed in \mulase. The polynomials now satisfy the equation
$$
P_3^2(z) +  \a\, z^2(z-1)^2 = Q_2^3(z) \,.
$$
Solving the differential equation \difftwo\ for this case, we find the unique solution
$$\eqalign{
P_3(z) = 1-{3\over 2}z -{3\over 2}z^2 + z^3 \qquad
Q_2(z) = 1-z+z^2 \quad \hbox{and} \qquad \a =  {3 \sqrt{3}\over 2}
}$$
and substituting this into \belyimap\ we get
$$
\b(z) = {4\over 27} {(z^2-z+1)^3 \over z^2(z-1)^2} \,,
$$
which is the answer quoted in \mulase.

\subsec{Some Comments on Belyi Maps, Children's Drawings and Seiberg-Witten Theory}

It turns out that the surprising connection we have found between Belyi maps and Argyres-Douglas curves (in \qeqi) can be generalized to any Seiberg-Witten curve that develops an isolated singularity. The precise connection will be explored in a forthcoming publication \AshokCD. Here we just collect a few well known facts about Belyi maps and comment briefly on the connection. The proofs of most of the technical statements to follow can be found in \leila$\,$.

If we are given a Belyi map in terms of rational functions, say
\eqn\generalB{
\b(z) = {A(z) \over B(z)} \,,
}
it is always possible to associate to it a diagram on the Riemann sphere. As explained in \refs{\leila, \mulase}, these graphs (usually referred to as Grothendieck's Dessins d'Enfants or ``children's drawings") are given by the inverse image under $\b$ of the interval $[0,1]$ on $\IP^1\,$:
\eqn\Dmap{
D = \b^{-1}([0,1]) \,.
}

A dessin is said to be ``clean" if all the ramification indices at the pre-images of $1$ are exactly equal to two. We see that this is satisfied by \belyimap\ and so the dessins we construct using \Dmap\ are clean. More generally, the Belyi map in \generalB\ leads to clean dessins iff the polynomials in \generalB\  satisfy the equation
\eqn\generalpoly{
A(z)-B(z) = P^2(z) \,,
}
for some $P(z)$. It can be shown that a clean  dessin has as many vertices as there are distinct zeroes of the Belyi map and as many open cells as there are distinct poles. The valency of each vertex is given by the order of the corresponding zero. Similarly, the valency of each open cell (i.e. the number of edges that bound it) is given by the order of the pole. 

It is not difficult to see from these definitions that, when the polynomial equation \generalpoly\ describes the Argyres-Douglas singularity in \qeqi$\,$, the dessin defined by \Dmap\  coincides with the critical graph of the Strebel differential. More generally, we can identify any polynomial equation \generalpoly\ that gives rise to a clean Belyi map with a Seiberg-Witten curve. From \Dmap\ above, there will be a children's drawing associated with any such Seiberg-Witten curve. An example that will be explored in detail in this direction  \AshokCD\ is that of pure $U(N)$ gauge theory, for which $B(z) = -4\Lambda^{2N}$, with $\Lambda$ the strong coupling scale of the theory.

\subsec{Assigning Lengths To Drawings}

From our discussion at the end of Section $3.1$, it is clear that the Strebel differentials coming from Belyi maps have integral lengths. We now extend this by associating a general differential $\phi_{D}$ to any drawing $D$ that can be obtained from a Belyi map using \Dmap. Using $\phi_{D}$ one can assign lengths to the edges of the drawing. Let us see how this comes about. 

It follows from the definition of the Belyi map in \generalB\ that every edge which goes between any two successive zeroes of $A(z)$ (pre-image of $0$) has on it a pre-image of $1$, which corresponds to a zero of $P(z)$. This is shown below in Figure $6$ for the specific case $A(z)=Q_4^3(z)$, $B(z)=B_3^3(z)$ and $P(z)$ a degree $6$ polynomial: a case discussed in Section $5$ as the $(3,3,3,3)$ example. 
\bigskip
\centerline{\epsfxsize=0.70\hsize\epsfbox{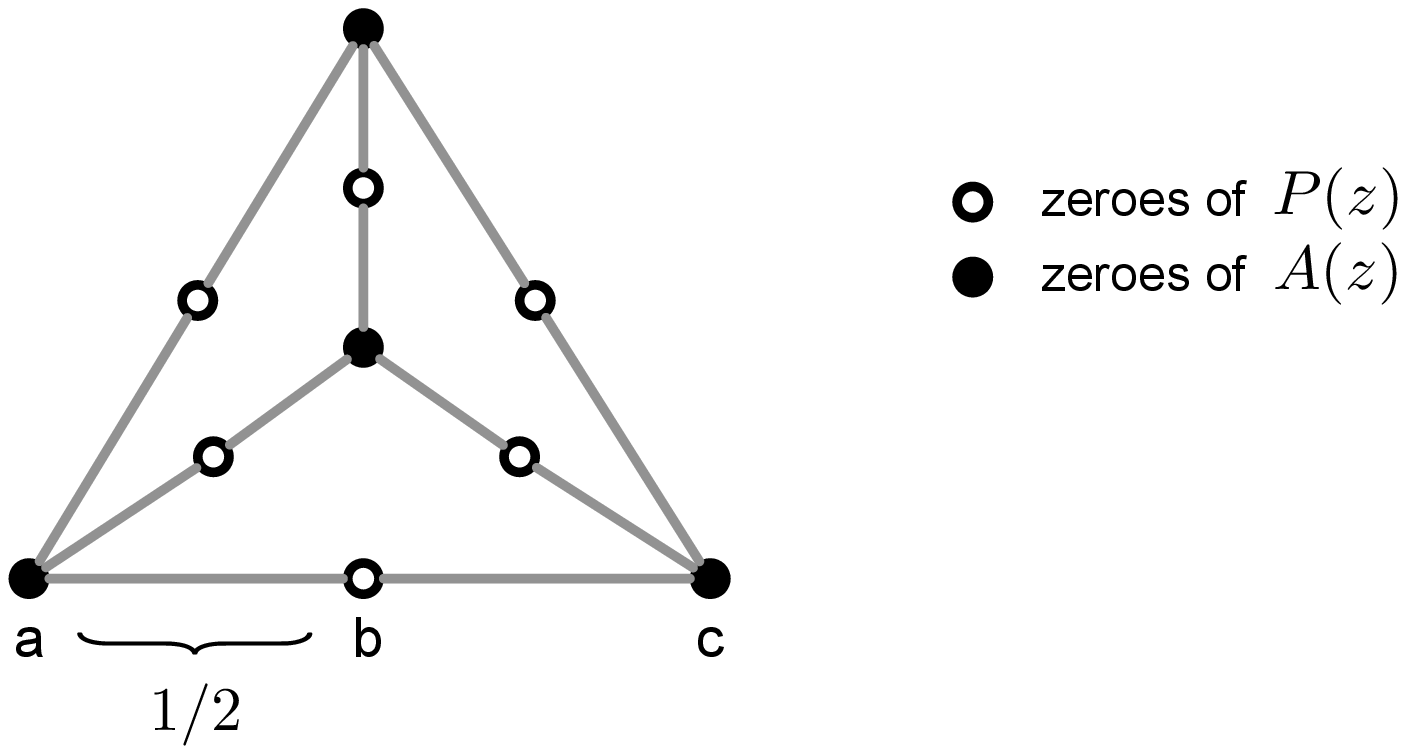}}
\noindent{\ninepoint\sl \baselineskip=3pt {\bf Figure $6$}:{\sl $\;$
Location of the pre-images of $0$ and $1$ under the Belyi map$\,$. Pre-images of $0$ correspond to zeroes of $A(z)$ and pre-images of $1$ correspond to zeroes of $P(z)$.}}
\medskip

Let us use the definitions in \phitheta\ and \thetaz\ for polynomials $P(z)$ and $B(z)$ that satisfy \generalpoly\ to construct the differential $\phi_{D}\,$: 
\eqn\defH{
\sqrt{\phi_{D}(z)}\, dz = {1\over 2\pi i}\,  d\log\left({P(z)-\sqrt{A(z)} \over P(z)+\sqrt{A(z)}}\right) \,.
}
From \defH\ and \phitheta\ it follows that
\eqn\lengtwo{
{1\over 2\pi i} \int_{a}^{b}\, \sqrt{\phi_{D}(z)}\, dz = {1\over 2\pi} \int_{\theta_a}^{\theta_b}\, d\theta \,,
}
where $a$ and $b$ are the zeroes of $A(z)$ and $P(z)$ respectively and $\theta_a$ and $\theta_b$ are the corresponding $\theta$-coordinates. From \defH\ we see that at a zero of $A(z)$,  we get $\theta_a=0\ \hbox{mod}\ 2\pi$ and at a zero of $P(z)$, we get $\theta_b=\pi\ \hbox{mod}\ 2\pi$. Substituting these values into \lengtwo\ leads to 
\eqn\lengthgen{
{1\over 2\pi i} \int_{a}^{b}\, \sqrt{\phi_{D}(z)}\, dz = \half + \Bbb{Z}\,.
}
As shown in Figure $6$, when $a$ and $b$ are adjacent on the graph, we get 
\eqn\length{
{1\over 2\pi i} \int_{a}^{b}\, \sqrt{\phi_{D}(z)}\, dz = \half \,.
}

\newsec{Concluding Remarks}

We have shown that finding Strebel differentials with integer lengths is equivalent to solving an algebraic problem. This algebraic problem is the same as that of finding generalized Argyres-Douglas singularities in the Coulomb moduli space of an $\CN=2$ gauge theory, which corresponds to a factorization of the Seiberg-Witten curve of the form \qeqi.

This correspondence between Strebel differentials and $\CN=2$ gauge
theory turns out to be quite useful: the relation to Strebel
differentials allowed us to write down the differential equation \difftwo\ that
was instrumental in solving the factorization problem.

We showed, using the correspondence of Strebel differentials to
metric ribbon graphs, that any Strebel differential can be
approximated by one with integer lengths divided by an appropriate
integer. In particular, for the example with four punctures, by solving the $(n,n,n,n)$
case with integer lengths we could obtain the set of
points in the moduli space with redidues $(1,1,1,1)$ that lead to
rational Strebel lengths. One can see from Figure $4$ that the
lattice of points that solve the $(n,n,n,n)$ problem get finer and
finer as one solves for higher and higher $n$ values. Taking $n$ to
be very large, one can intuitively understand how the solutions the
Strebel problem with rational lengths form a dense subset of the
moduli space. It would be interesting to develop a method for finding
an interpolating function $q_1=q_1(t)$ from the data we have.

The correspondence between Strebel differentials and metric ribbon
graphs also allowed us to find the number of solutions to the
factorization problem \qeqi, something that is not at all obvious
from the gauge theory side. For all residues equal to $n$, we found
a particularly simple answer: there are as many solutions as there
are partitions of $n$ into exactly three strictly positive integers.

The Strebel differential also has a simple interpretation in the
$\CN=2$ gauge theory: it is related to the generating function of
scalar correlation functions $T(z)\, dz$ as shown in \strebelSW. It has an even more
striking interpretation in string theory, in the context of
geometric engineering: it is the $H$-field integrated over an $S^2$
in the non-compact Calabi-Yau. It would be extremely interesting to
explore this in more detail. In particular, we found that the
conditions imposed on a general quadratic differential to be Strebel
were very similar to the conditions found in \ShapereV\ to find
special lagrangian three manifolds in a non-compact manifold.
Interestingly, the Calabi-Yau in question is conjectured to be
holographically dual to Argyres-Douglas superconformal theories \refs{\GiveonKO, \Pelc}.

Finally, from the discussion in Section $7$, it is clear that the relationship between Seiberg-Witten theories and clean Belyi maps is more general and not restricted to the special class of Argyres-Douglas singularities discussed in this paper. The relation between Belyi maps and the children's drawings raises the interesting question of understanding the role of such drawings in gauge theory. In this note we have been content with describing how our understanding of the relationship between Argyres-Douglas curves and Strebel differentials leads to a differential equation technique to solve the problem of factorizing the Seiberg-Witten curve as in \qeqi. We will explore the more intriguing relationship between children's drawings and Seiberg-Witten theory in a forthcoming publication \AshokCD.

\medskip

\centerline{\bf Acknowledgments}

\medskip

The authors would like to thank Camille Boucher-Veronneau for early participation in the project. We would also like to thank Philip Argyres, Paolo Benincasa, Alex Buchel, Justin David, Jaume Gomis and Rajesh Gopakumar for discussions. The research of SA and FC at the Perimeter Institute is supported in part by funds from NSERC of Canada and MEDT of Ontario. The research of ED is supported in part by the National Science Foundation under Grant No. PHY99-07949.

\appendix{A}{Maximally Confining Vacua}

The maximally confining vacua \refs{\SeibergWone, \DouglasS} are isolated singular points of the Seiberg-Witten curve of pure $\N=2$ $SU(N)$ SYM which correspond to the factorization problem
\eqn\fofo{P^2_N(z) - 4 = (z^2-4)H^2_{N-1}(z)\,.}
In this appendix we show how differentiation gives a simple way of
finding the solutions. After differentiating once on both sides we
get
\eqn\difa{2 P_N(z) P'_N(z) = H_{N-1}(z)\big[ 2z
H_{N-1}(z)+2(z^2-4)H'_{N-1}(z)\big]\,.}
Note that a root of  $P_N(z)$ on the l.h.s. cannot possibly be a
root of $H_{N-1}(z)$ on the r.h.s., because that would contradict
\fofo. Therefore, all roots of $H_{N-1}(z)$ must be roots of
$P'_N(z)$. Since the two polynomials $H_{N-1}(z)$ and $P_N(z)$ have the same degree and are both  monic, it must be that
\eqn\gagi{P'_N(z) = N H_{N-1}(z)\, .}
Taking this into account, equation \difa\ implies
\eqn\asu{ N P_N(z) = z H_{N-1}(z)+(z^2-4)H'_{N-1}(z)}
and using \gagi\ to solve for $H_{N-1}(z)$ we get
\eqn\zaza{N^2 P_N(z) = z {dP_N\over dz}(z) +
(z^2-4){d^2P_{N-1}\over dz^2}(z)\,.}
This differential equation is easily recognized as the Chebyshev equation and it has two independent solutions $T_N(z)$ and $\sqrt{z^2-4}\,U_{N-1}(z)\,$, where $T_N(z)$ and $U_{N-1}(z)$ are the Chebyshev polynomials of first and second kind respectively. Since $P_N(z)$ is a monic polynomial we conclude that
$$
P_N(z)= 2\, T_N\left({z\over 2}\right) \qquad \hbox{and}\qquad H_{N-1}(z) = U_{N-1}\left({z\over 2} \right) \,.
$$

\appendix{B}{Strebel Differential On A Sphere With Residues $(1,1,1,3n)$}

For this case one can solve the factorization problem \polyeq\  for any odd value of $n\,$. (For even $n$, one cannot define a polynomial factorization problem as the degree of $N$ is no longer integer.) Since we require the lengths to be integral, the only possible way to draw the critical graph -such that the sum of positive integers is equal to $1$ for three of the loops - is to have three of the edges be just curves that encircle the location of the poles at $(0,1,t)\,$. Since all vertices are trivalent, this fixes the form of the critical graph to be Figure $7\,$. The residue at infinity being $3n$ then leads to the condition
$$
\sum_{i=1}^{3} l_i = {3\over 2}(n-1) \,.
$$
As we argued in the main text, the number of solutions (distinct values of $t$ and $q_1$) of this problem is $p_3({3\over 2}(n-1))$, where $p_3(n)$ is the partition of $n$ into three non-zero positive integers. However, it is very easy to find {\it one} solution of the factorization problem: if we require that all $l_i$ be equal, then the points $(0,1,t)$ form an equilateral triangle, which fixes $t$ - and hence $q_1$ - to be
\eqn\ansatz{
t ={1\over 2 }+ i {\sqrt{3}\over 2}\qquad\qquad q_1=-2 -i {2\over \sqrt{3}} \,.
}
This completely determines $Q_4$ in \Qdefn. Moreover, since this case satisfies the condition that $m_{\infty} > m_0+m_1+m_t$, one can use \relas\ and solve for $P_N(z)$ completely. Substituting $P_N(z)$ and $Q_4(z)$ into the full differential equation \difftwo\ one can solve for $\a$. Plugging all this back into the original factorization problem, we can find $R(z)$.

\medskip
\centerline{\epsfxsize=0.60\hsize\epsfbox{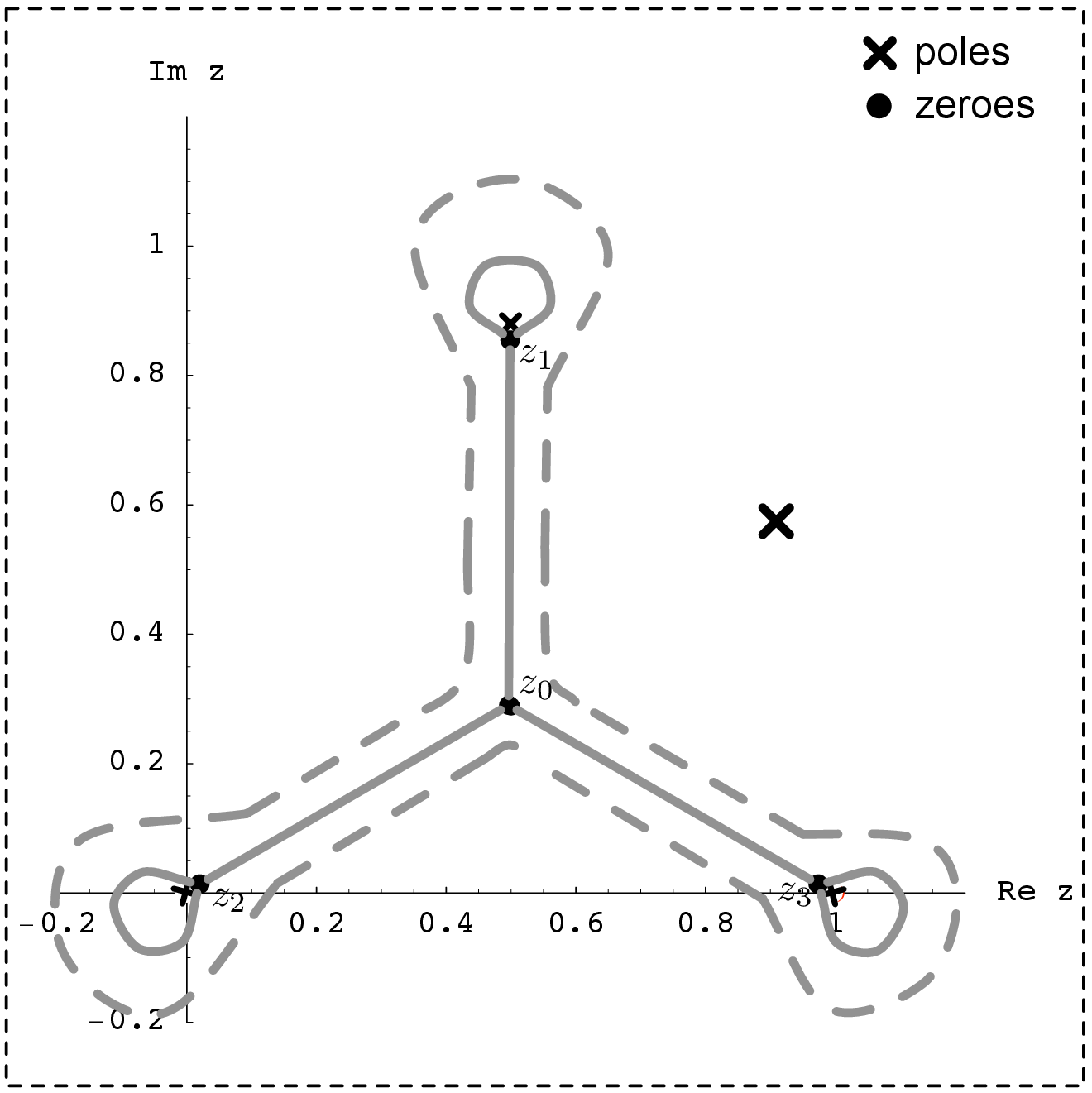}}
\noindent{\ninepoint\sl \baselineskip=3pt {\bf Figure $7$}:{\sl $\;$
Location of poles and zeroes of the Strebel differential with residues $(15,1,1,1)\,$. The dashed line around the critical graph represents the contour that computes the residue at infinity. }}
\medskip

In practice, this program is not easy to carry out. However, by solving the equations for the first few cases, we find that the factorization problem
$$
P^2_{{3\over 2}(n+1)}(z) + \a\, z(z-1)(z-t) = Q_4^3(z)\, R^2_{{3\over 2}(n-3)}(z) \,
$$
is solved by the ansatz \ansatz\ if $\a$ is given by the closed expression
$$
\alpha=  (-1)^{n+1\over 2}4i\, {({n+1\over 2})^{n+1}({n-1\over 2})^{n-1}\over 3^{3n\over 2} n^{2n}} \,.
$$
The Strebel differential for this problem has the form
$$
\phi(z) (dz)^2= -{9\,n^2 \over 4\pi^2 }\, {Q_4(z) \over z^2(z-1)^2(z-{1\over 2 }- i {\sqrt{3}\over 2})^2}\,(dz)^2\,,
$$
where
$$
Q_4(z)=z^4 +(-2-{2i\over\sqrt{3}})z^3+(1+i\sqrt{3})z^2-{i \over 3\sqrt{3} n^2}(1+3n^2) z+{1\over18 n^2}(-1+i \sqrt{3})\,.
$$
The zeroes of $Q_4(z)$ are located at
$$\eqalign{
z_1 &= {1 \over 6}(3+i\sqrt{3})\,,\quad z_2 = {1 \over 6}(3+i\sqrt{3})+{i \over 6 n^2}(3i+\sqrt{3})
(n^4-n^6)^{1\over 3} \cr
z_3 &= {1 \over 6}(3+i\sqrt{3}) -{i\over \sqrt{3}}(n^4-n^6)^{1\over 3}\,,\quad z_4 = {1 \over 6}(3+i\sqrt{3})\left(1+ {(n^4-n^6)^{1\over 3} \over n^2}\right)\,.
}$$
One can explicitly compute the Strebel lengths and check that they are ${n-1\over 2}$ for $|z_{0i}|$, $i=1,2,3$. The notation is as in Figure $7\,$, where we have shown the critical graph for the particular case $n=5\,$. For this the edges have lengths $z_{0i}= 2$ and $z_{ii}=1$ so the residue at infinity is (correctly) given by $m_{\infty} = 15$.

\listrefs
\end